# Rankers over Infinite Words*

Luc Dartois[1] Manfred Kufleitner[2] Alexander Lauser[2]

[1] ENS Cachan, France
ldartois@dptinfo.ens-cachan.fr

[2] FMI, Universität Stuttgart, Germany
{kufleitner, lauser}@fmi.uni-stuttgart.de

May 3, 2010

**Abstract.** We consider the fragments $\mathrm{FO}^2$, $\Sigma_2 \cap \mathrm{FO}^2$, $\Pi_2 \cap \mathrm{FO}^2$, and $\Delta_2$ of first-order logic $\mathrm{FO}[<]$ over finite and infinite words. For all four fragments, we give characterizations in terms of rankers. In particular, we generalize the notion of a ranker to infinite words in two possible ways. Both extensions are natural in the sense that over finite words, they coincide with classical rankers and over infinite words, they both have the full expressive power of $\mathrm{FO}^2$. Moreover, the first extension of rankers admits a characterization of $\Sigma_2 \cap \mathrm{FO}^2$ while the other leads to a characterization of $\Pi_2 \cap \mathrm{FO}^2$. Both versions of rankers yield characterizations of the fragment $\Delta_2 = \Sigma_2 \cap \Pi_2$. As a byproduct, we also obtain characterizations based on unambiguous temporal logic and unambiguous interval temporal logic.

## 1 Introduction

We consider fragments of two-variable first-order logic $\mathrm{FO}^2$. Formulas are interpreted over words which may be infinite or finite. Over finite words only, a large number of different characterizations of $\mathrm{FO}^2$ is known, see e.g. [7] or [1] for an overview. Some of the characterizations have been generalized to infinite words in [2]. In this paper, we continue this line of work. For this paper the main difference between finite word models and infinite word models is the following: Over finite words, $\mathrm{FO}^2$ and the fragment $\Delta_2 = \Sigma_2 \cap \Pi_2$ have the same expressive power [8], whereas $\Delta_2$ is a strict subclass of $\mathrm{FO}^2$ over infinite words. Moreover, in the case of infinite words, $\mathrm{FO}^2$ is incomparable to $\Sigma_2$ and $\Pi_2$. By definition $\Sigma_2$ is the class of formulas in prenex normal form with two blocks of quantifiers starting with a block of existential quantifiers, and $\Pi_2$ is the class of negations of $\Sigma_2$-formulas. Here and throughout the paper, we identify a logical fragment with the class of languages definable in the fragment.

An important concept in this paper are rankers which have been introduced by Immerman and Weis [9] in order to give a combinatorial characterization of quantifier alternation within $\mathrm{FO}^2$ over finite words. Casually speaking, a *ranker* is a sequence of instructions of the form *"go to the next a-position"* and *"go to the previous a-position"* for some letters $a$. For every word, a ranker is either undefined or it determines a unique position. We generalize rankers to infinite words in two

*The last two authors acknowledge the support by the German Research Foundation (DFG) for the project DI 435/5-1.

possible ways. The main difference to finite words is that we have to define the semantics of *"go to the last a-position"* if there are infinitely many occurrences of the letter $a$. The first solution is to say that this modality evaluates to false and that the position is undefined. The second approach is to stay at an infinite position. For example, if a word has infinitely many $a$-positions but only two $b$-positions, then in the first semantics *"go to the last a-position and from there, go to the previous b-position"* would be false while in the second semantics it would be true and it would determine the last $b$-position. By delaying the interpretation of modalities until some letter with finite occurrence is met, the second semantics is reminiscent of the *lazy evaluation* principle. We therefore call the second semantics *lazy rankers*. If we want to emphasize that we use the first semantics, then we often use the term *eager ranker*. The language $L(r)$ of a ranker $r$ consists of all words such that $r$ is defined. A *ranker language* is a Boolean combination of languages of the form $L(r)$ for some rankers $r$.

In both ways, rankers admit natural combinatorial characterizations of the first-order fragments $\mathrm{FO}^2$ and $\Delta_2$ over finite and infinite words. Moreover, the eager semantics yields a characterization of $\Sigma_2 \cap \mathrm{FO}^2$ while lazy rankers lead to a characterization of $\Pi_2 \cap \mathrm{FO}^2$. We note that the decidability results for the first-order fragments lead to decidability results for the respective ranker fragments [2].

Let $\Gamma^\infty$ be the set of all finite and infinite words over the alphabet $\Gamma$ and let $L \subseteq \Gamma^\infty$. Our main results are

- $L \in \mathrm{FO}^2$ if and only if $L$ is an eager ranker language (Theorem 1) if and only if $L$ is a lazy ranker language (Theorem 5).
- $L \in \Sigma_2 \cap \mathrm{FO}^2$ if and only if $L$ is a positive eager ranker language with some additional atomic modality (Theorem 2).
- $L \in \Pi_2 \cap \mathrm{FO}^2$ if and only if $L$ is a positive lazy ranker language with some additional atomic modality (Theorem 4).
- $L \in \Delta_2$ if and only if $L$ is a ranker language such that all instructions are starting with a modality *"go to the first a-position"* (Theorem 3).

It turns out that unambiguous temporal logic [3] and unambiguous interval temporal logic [4] allow natural intermediate characterizations on the way from first-order logic to rankers. In particular, this yields temporal logic counterparts of the first-order fragments. Moreover, we present a way for converting formulas in unambiguous interval temporal logic into equivalent formulas in unambiguous temporal logic, which does not introduce new negations (Propositions 1 and 2). This also leads to a new characterization of $\mathrm{FO}^2$ over finite words in terms of restricted ranker languages (Corollary 1).

In this paper, all steps from fragments of first-order logic to interval temporal logic are based on characterizations in terms of so-called unambiguous polynomials; almost all other steps are effective syntactic transformations. The sole exception is the inclusion of some ranker fragment in $\Pi_2 \cap \mathrm{FO}^2$. This step relies on a characterization of $\Pi_2 \cap \mathrm{FO}^2$ in terms of the alphabetic topology [2].

An extended abstract of our results will be presented at the 14th International Conference on Developments in Language Theory (DLT 2010).

## 2 Preliminaries

In the following $\Gamma$ denotes a finite alphabet. For $A \subseteq \Gamma$ we denote by $A^*$ the set of finite words over $A$. The set of infinite words is $A^\omega$, and $A^\infty = A^* \cup A^\omega$ is the set of finite and infinite words. The empty word is $\varepsilon$ and we have $\{\varepsilon\} = \emptyset^\infty$. We denote potentially infinite words by lowercase

Greek letters $\alpha, \beta, \gamma$ whereas finite words are denoted by lowercase Latin letters $u, v, w$; for letters in $\Gamma$ we use $a, b, c, d$. For a word $\alpha$ and a position $x$ of the word, $\alpha(x)$ is the $x$-th letter of $\alpha$. By $|\alpha| \in \mathbb{N} \cup \{\infty\}$ we denote the *length* of $\alpha$. Therefore $\alpha = \alpha(1) \cdots \alpha(|\alpha|)$ if $\alpha$ is finite and $\alpha = \alpha(1)\alpha(2) \cdots$ if $\alpha$ is infinite. We extend this notation to intervals $T \subseteq \mathbb{N}$ and write $\alpha(T)$ for the word comprising the positions of $\alpha$ contained in $T$. In particular, we do not require that $T$ is contained in the set of positions of $\alpha$. Hence, for all $\alpha \in \Gamma^\infty$ we have $\alpha(\emptyset) = \varepsilon$, and $\alpha(\mathbb{N}) = \alpha$ even if the word $\alpha$ is finite. We call $\mathrm{alph}(\alpha)$ the *alphabet* of $\alpha$, i.e., the set of letters occurring in $\alpha$. For $a \in \Gamma$, a position labeled by $a$ is called an *$a$-position.* By $\mathrm{im}(\alpha)$ we mean the *imaginary* alphabet of $\alpha$, i.e., the set of letters occurring infinitely often in $\alpha$. For $A \subseteq \Gamma$, the set of words with imaginary alphabet $A$ is denoted by $A^{\mathrm{im}}$. In particular, $\Gamma^* = \emptyset^{\mathrm{im}}$.

A *monomial* (of *degree* $k$) is a language of the form $A_1^* a_1 \cdots A_k^* a_k A_{k+1}^\infty$ for letters $a_i \in \Gamma$ and sets $A_i \subseteq \Gamma$. It is *unambiguous* if each word of the monomial has a unique factorization $u_1 a_1 \cdots u_k a_k \beta$ with $u_i \in A_i^*$ and $\beta \in A_{k+1}^\infty$. A *polynomial* (of *degree* $k$) is a finite union of monomials (of degree at most $k$). It is called *unambiguous* if it is a finite union of unambiguous monomials.

**Example 1** The set of all finite words over an alphabet $A \subseteq \Gamma$ is an unambiguous polynomial. We have

$$A^* \;=\; \emptyset^\infty \;\cup\; \bigcup_{a \in A} A^* a \emptyset^\infty,$$

i.e., a word is of finite length if it is either empty or if there is a last letter. ◊

**Example 2** Consider the language $L = (\Gamma \setminus \{b\})^* a \Gamma^\infty \;\cap\; (\Gamma \setminus \{c\})^* b \Gamma^\infty \;\cap\; \Gamma^* c \Gamma^\infty$ of all words containing a $c$ such that there is an $a$ with no $b$ to the left, and such that there is a $b$ with no $c$ to the left. This language is an unambiguous monomial since

$$L = (\Gamma \setminus \{b, c\})^* a (\Gamma \setminus \{c\})^* b \Gamma^* c \Gamma^\infty.$$

Moreover, $L$ is the set of all words such that the first $a$ occurs before the first $b$ which in turn occurs before the first $c$. ◊

## 2.1 Fragments of First-Order Logic

We denote by $\mathrm{FO} = \mathrm{FO}[<]$ the first-order logic over words interpreted as labeled linear orders (without $\infty$). As atomic formulas, FO comprises $\top$ (for *true*), the unary predicate $\lambda(x) = a$ for $a \in \Gamma$, and the binary predicate $x < y$ for variables $x$ and $y$. The idea is that variables range over the linearly ordered positions of a word and $\lambda(x) = a$ means that $x$ is an $a$-position. Apart from the Boolean connectives, we allow composition of formulas using existential quantification $\exists x \colon \varphi$ and universal quantification $\forall x \colon \varphi$ for $\varphi \in \mathrm{FO}$. The semantics is as usual. We introduce the common shortcut $\bot$ for $\neg\top$. Typical names for formulas in this paper are $\varphi, \psi, \varrho, \vartheta, \mu, \nu, \sigma$.

Every formula in FO can be converted into a semantically equivalent formula in prenex normal form by renaming variables and moving quantifiers to the front. This observation gives rise to the fragment $\Sigma_2$ (resp. $\Pi_2$) consisting of all FO-formulas in prenex normal form with only two blocks of quantifiers, starting with a block of existential quantifiers (resp. universal quantifiers). Note that the negation of a formula in $\Sigma_2$ is equivalent to a formula in $\Pi_2$ and vice versa. The fragments $\Sigma_2$ and $\Pi_2$ are both closed under conjunction and disjunction. Furthermore, $\mathrm{FO}^2$ is the fragment of FO containing all formulas which use at most two different names for the variables. This is a natural restriction, since FO with three variables already has the full expressive power of FO.

A *sentence* in FO is a formula without free variables. For a sentence $\varphi$ the *language defined by* $\varphi$, denoted by $L(\varphi)$, is the set of all words $\alpha \in \Gamma^\infty$ that model $\varphi$, i.e., $\alpha \models \varphi$. We frequently identify logical fragments with the classes of languages they define (as in the definition of the fragment $\Delta_2 = \Sigma_2 \cap \Pi_2$ for example).

**Example 3** Consider the formulas

$$\varphi \;=\; \exists x \forall y \colon y \leq x \vee \lambda(y) \neq a \qquad \text{and} \qquad \psi \;=\; \forall x \exists y \colon y > x \wedge \lambda(y) = a.$$

The formula $\varphi \in \Sigma_2 \cap \mathrm{FO}^2$ states that after some position there is no $a$-position, i.e., $L(\varphi)$ contains all words with finitely many $a$-positions. Its negation $\psi \in \mathrm{FO}^2 \cap \Pi^2$ says that for all positions there is a greater $a$-position, i.e., $L(\psi)$ is set of all words $\alpha$ with $a \in \mathrm{im}(\alpha)$. Surprisingly, $L(\varphi)$ is not definable in $\Pi_2$ and $L(\psi)$ is not definable in $\Sigma_2$, cf. [2].

# 3 Rankers and Unambiguous Temporal Logics

For finite words, rankers have been introduced by Immerman and Weis [9]. They can be seen as a generalization of *turtle programs* used by Schwentick, Thérien, and Vollmer [6] for characterizing $\mathrm{FO}^2$-definable languages over finite words. The main difference between rankers and turtle programs is that rankers either uniquely determine a position in a word or they are undefined, whereas turtle programs mainly distinguish between being defined and being undefined.

Extending rankers with Boolean connectives yields unambiguous temporal logic (unambiguous TL). It is called *unambiguous* since each position considered by some formula in this logic is unique. Unambiguous TL has been introduced for Mazurkiewicz traces [3] which are a generalization of finite words.

All of our characterizations of first-order fragments rely on unambiguous polynomials. A natural intermediate step from polynomials to temporal logic is interval temporal logic. Unambiguous interval temporal logic (unambiguous ITL) has been introduced by Lodaya, Pandya, and Shah [4] for finite words. They showed that over finite words it has the same expressive power as $\mathrm{FO}^2$.

In this section, we generalize all three concepts (rankers, unambiguous TL, and unambiguous ITL) to infinite words. For each concept there are essentially two natural choices for such generalizations. Surprisingly, it turns out that one extension can be used for the characterization of the first-order fragment $\Sigma_2 \cap \mathrm{FO}^2$ over $\Gamma^\infty$ while the other yields a characterization of $\Pi_2 \cap \mathrm{FO}^2$. Moreover, both semantics can be used for describing $\mathrm{FO}^2$ and $\Delta_2$. In fact, for $\Delta_2$ we use some fragment of rankers which conceals the difference between the two versions.

Basically, all proofs of our main theorems have the following structure: Using some characterization in terms of unambiguous polynomials, we go from first-order logic to interval temporal logic; then formulas in interval temporal logic are transformed into equivalent formulas in temporal logic, which in turn can be easily converted into some ranker descriptions. The last step is to express ranker languages within some fragment of first-order logic. In all proofs, the main technical step is the conversion of unambiguous ITL into unambiguous TL without introducing new negations (Propositions 1 and 2).

## 3.1 Rankers

A *ranker* is a finite word over the alphabet $\{\mathsf{X}_a, \mathsf{Y}_a \mid a \in \Gamma\}$. It can be interpreted as a sequence of instructions of the form $\mathsf{X}_a$ and $\mathsf{Y}_a$. Here, $\mathsf{X}_a$ (for neXt-$a$) means "go to the next $a$-position" and $\mathsf{Y}_a$ (for Yesterday-$a$) means "go to the previous $a$-position". Below, we will introduce a second variant of rankers (lazy rankers). For distinguishing, we will sometimes use the attribute *eager* for this first version of rankers. For a word $\alpha$ and a position $x \in \mathbb{N} \cup \{\infty\}$ we define

$$\begin{aligned} \mathsf{X}_a(\alpha, x) &= \min \{ y \in \mathbb{N} \mid \alpha(y) = a \text{ and } y > x \}, \\ \mathsf{Y}_a(\alpha, x) &= \max \{ y \in \mathbb{N} \mid \alpha(y) = a \text{ and } y < x \}. \end{aligned}$$

As usual, we set $y < \infty$ for all $y \in \mathbb{N}$. The minimum and the maximum of $\emptyset$ as well as the maximum of an infinite set are undefined. In particular, $\mathsf{X}_a(\alpha, \infty)$ is always undefined and $\mathsf{Y}_a(\alpha, \infty)$ is defined if and only if $a \in \mathrm{alph}(\alpha) \setminus \mathrm{im}(\alpha)$. We extend this definition to rankers by setting $\mathsf{X}_a\, r(\alpha, x) = r(\alpha, \mathsf{X}_a(\alpha, x))$ and $\mathsf{Y}_a\, r(\alpha, x) = r(\alpha, \mathsf{Y}_a(\alpha, x))$, i.e., rankers are processed from left to right. We say that $r(\alpha, x)$ is undefined, if after processing some prefix of $r$ on $\alpha$, the resulting position is undefined. If $r(\alpha, x)$ is defined for some non-empty ranker $r$, then $r(\alpha, x) \neq \infty$.

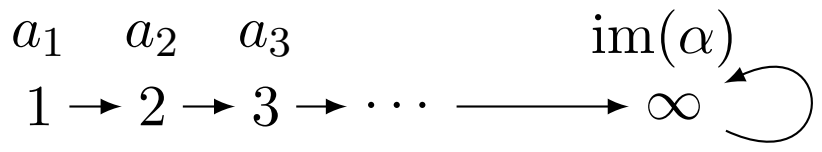

Figure 1: Signature of $\alpha = a_1\, a_2\, a_3 \cdots$ over lazy rankers

Next, we define another variant of rankers as finite words over the alphabet $\{\mathsf{X}^\ell_a, \mathsf{Y}^\ell_a \mid a \in \Gamma\}$. The superscript $\ell$ is derived from *lazy* and accordingly such rankers are called *lazy rankers*. The difference to eager rankers is that lazy rankers can point to an infinite position $\infty$. The idea is that the position $\infty$ is not reachable from any finite position and that it represents the behavior at infinity. We imagine that $\infty$ is greater than all finite positions and it is labeled by all letters in $\mathrm{im}(\alpha)$ for words $\alpha$. Therefore, it is often adequate to set $\infty < \infty$, since the infinite position simulates a set of finite positions. For a word $\alpha$ and a finite position $x \in \mathbb{N}$ we define $\mathsf{X}^\ell_a(\alpha, x) = \mathsf{X}_a(\alpha, x)$ and $\mathsf{Y}^\ell_a(\alpha, x) = \mathsf{Y}_a(\alpha, x)$. For the infinite position we set

$$\mathsf{X}^\ell_a(\alpha, \infty) = \begin{cases} \infty & \text{if } a \in \mathrm{im}(\alpha) \\ \textit{undefined} & \text{else} \end{cases}$$

$$\mathsf{Y}^\ell_a(\alpha, \infty) = \begin{cases} \infty & \text{if } a \in \mathrm{im}(\alpha) \\ \mathsf{Y}_a(\alpha, \infty) & \text{else} \end{cases}$$

i.e., $\mathsf{Y}^\ell_a(\alpha, \infty)$ is undefined if $a \notin \mathrm{alph}(\alpha)$, and $\mathsf{Y}^\ell_a(\alpha, \infty) = \mathsf{Y}_a(\alpha, \infty)$ is a finite position if $a \in \mathrm{alph}(\alpha) \setminus \mathrm{im}(\alpha)$. As before, we extend this definition to rankers of length $> 1$ by setting $\mathsf{X}^\ell_a\, r(\alpha, x) = r(\alpha, \mathsf{X}^\ell_a(\alpha, x))$ and $\mathsf{Y}^\ell_a\, r(\alpha, x) = r(\alpha, \mathsf{Y}^\ell_a(\alpha, x))$. We denote by

$$\mathrm{alph}_\Gamma(r) = \{a \in \Gamma \mid r \in q\, \{\mathsf{X}_a, \mathsf{Y}_a, \mathsf{X}^\ell_a, \mathsf{Y}^\ell_a\}\, s \text{ for some rankers } q, s\}$$

the set of letters in $\Gamma$ occurring in some modality of the ranker $r$. It can happen that $r(\alpha, \infty) = \infty$ for some non-empty lazy ranker $r$. This is the case if and only if $r$ is of the form $\mathsf{Y}^\ell_a\, s$ and $\mathrm{alph}_\Gamma(r) \subseteq \mathrm{im}(\alpha)$.

If the reference to the word $\alpha$ is clear from the context, then for eager and lazy rankers $r$ we shorten the notation and write $r(x)$ instead of $r(\alpha, x)$.

An eager ranker $r$ is an $\mathsf{X}$-*ranker* if $r = \mathsf{X}_a\, s$ for some ranker $s$ and $a \in \Gamma$, and it is a $\mathsf{Y}$-*ranker* if $r$ is of the form $\mathsf{Y}_a\, s$. Lazy $\mathsf{X}^\ell$-rankers and $\mathsf{Y}^\ell$-rankers are defined similarly. We proceed to define $r(\alpha)$, the position of $\alpha$ reached by the ranker $r$ by starting "outside" the word $\alpha$. The position $r(\alpha)$ is called $r$-*position*. The intuition is as follows. If $r$ is an $\mathsf{X}$-ranker or an $\mathsf{X}^\ell$-ranker, we imagine that we start at an outside position in front of $\alpha$; if $r$ is a $\mathsf{Y}$-ranker or a $\mathsf{Y}^\ell$-ranker, then we start at a position behind $\alpha$. Therefore, we define

$$\begin{aligned} r(\alpha) &= r(\alpha, 0) && \text{if } r \text{ is an } \mathsf{X}\text{-ranker or an } \mathsf{X}^\ell\text{-ranker}, \\ r(\alpha) &= r(\alpha, \infty) && \text{if } r \text{ is a } \mathsf{Y}\text{-ranker or a } \mathsf{Y}^\ell\text{-ranker}. \end{aligned}$$

On the left side of Figure 2, a possible situation for the eager ranker $\mathsf{Y}_a\, \mathsf{Y}_b\, \mathsf{X}_c$ being defined on some word $\alpha$ is depicted. The right side of the same figure illustrates a similar situation for the lazy ranker $\mathsf{Y}^\ell_d\, \mathsf{X}^\ell_d\, \mathsf{Y}^\ell_a\, \mathsf{Y}^\ell_b\, \mathsf{X}^\ell_c$ with $d \in \mathrm{im}(\alpha)$ and $a \in \mathrm{alph}(\alpha) \setminus \mathrm{im}(\alpha)$. Note that the eager version of the same ranker would not be defined on $\alpha$ since $d \in \mathrm{im}(\alpha)$.

Given these definitions, an eager ranker can either be undefined on a word $\alpha$ (if at some state of the evaluation of $r(\alpha)$ an instruction cannot be accomplished) or the ranker is defined on $\alpha$ and in this case it determines a unique finite position of $\alpha$. If a lazy ranker $r$ is defined on $\alpha$, then either $r(\alpha) = \infty$ or it defines a unique finite position of $\alpha$. For the empty ranker $\varepsilon$ we have $\varepsilon\,\mathsf{X}_a = \mathsf{X}_a$ and $\varepsilon\,\mathsf{X}^\ell_a = \mathsf{X}^\ell_a$ as well as $\varepsilon\,\mathsf{Y}_a = \mathsf{Y}_a$ and $\varepsilon\,\mathsf{Y}^\ell_a = \mathsf{Y}^\ell_a$, i.e., the empty ranker $\varepsilon$ either starts at position 0 or $\infty$ depending on whether the next modality is in $\{\mathsf{X}_a, \mathsf{X}^\ell_a \mid a \in \Gamma\}$ or in $\{\mathsf{Y}_a, \mathsf{Y}^\ell_a \mid a \in \Gamma\}$. Moreover, $\varepsilon$ is defined on every word even though it does not determine a unique position of the word. The empty ranker is both eager and lazy.

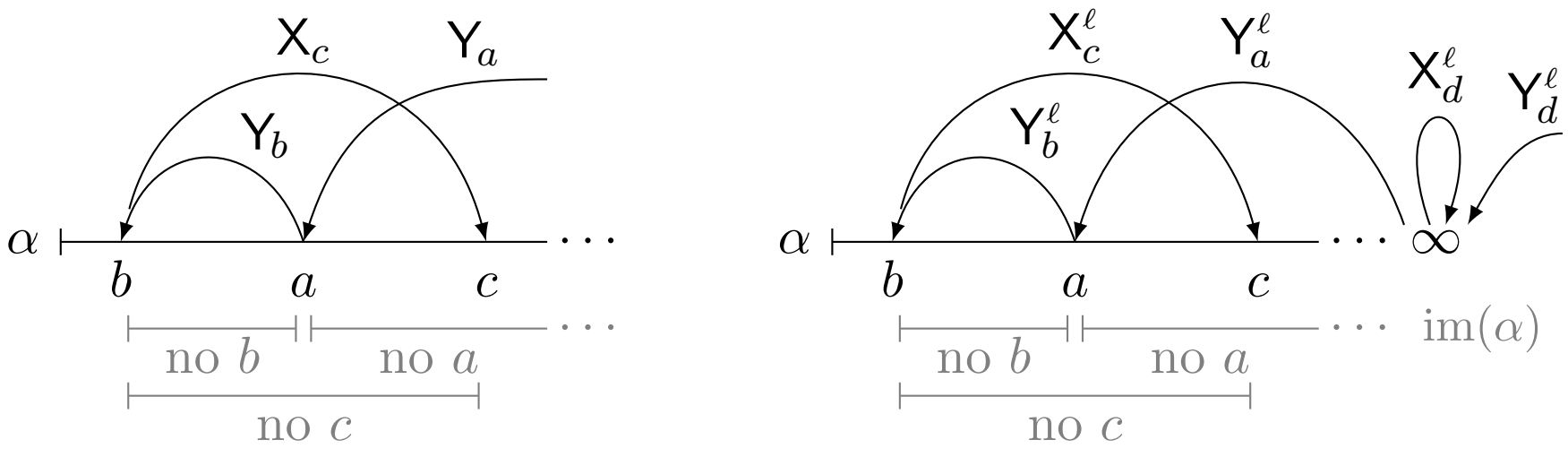

Figure 2: Eager and lazy rankers

For an eager or lazy ranker $r$ the language $L(r)$ generated by $r$ is the set of all words in $\Gamma^\infty$ on which $r$ is defined. A *(positive) ranker language* is a finite (positive) Boolean combination of languages of the form $L(r)$ for a ranker $r$. A *(positive) lazy ranker language* is a finite (positive) Boolean combination of languages of the form $L(r)$ for a lazy ranker $r$. Finally, a *(positive)* $\mathsf{X}$*-ranker language* is a (positive) ranker language using only $\mathsf{X}$-rankers.

**Example 4** The language $L = (\Gamma \setminus \{a,b,c\})^* a (\Gamma \setminus \{b,c\})^* b (\Gamma \setminus \{c\})^* c \Gamma^\infty$ of all words such that the first $a$ occurs before the first $b$ which in turn occurs before the first $c$ is a positive $\mathsf{X}$-ranker language because

$$L \;=\; L(\mathsf{X}_b\,\mathsf{Y}_a) \;\cap\; L(\mathsf{X}_c\,\mathsf{Y}_b).$$

We have $\mathrm{alph}_\Gamma(\mathsf{X}_b\,\mathsf{Y}_a) = \{a,b\}$ and $\mathrm{alph}_\Gamma(\mathsf{X}_c\,\mathsf{Y}_b) = \{b,c\}$, and both rankers are $\mathsf{X}$-rankers. $\Diamond$

**Example 5** Consider the language $L \subseteq \Gamma^\infty$ consisting of all non-empty words with $a$ as the first letter. A word is contained in $L$ if and only if it contains an $a$-position and such that no $b \in \Gamma$ occurs to the left of the first $a$-position. Therefore,

$$L \;=\; L(\mathsf{X}_a) \;\cap\; \Gamma^\infty \setminus \bigcup_{b \in \Gamma} L(\mathsf{X}_a\,\mathsf{Y}_b),$$

that is, $L$ is a Boolean combination of the rankers $\mathsf{X}_a$ and $\mathsf{X}_a\,\mathsf{Y}_b$ for $b \in \Gamma$. $\Diamond$

## 3.2 Unambiguous Temporal Logic

Our generalization of rankers allows us to define unambiguous temporal logic (unambiguous TL) over infinite words. As for rankers, we have an eager and a lazy variant. The syntax is given by:

$$\top \mid \neg\varphi \mid \varphi \vee \psi \mid \varphi \wedge \psi \mid \mathsf{X}_a\,\varphi \mid \mathsf{Y}_a\,\varphi \mid \mathsf{G}_{\bar a} \mid \mathsf{H}_{\bar a} \mid \mathsf{X}^\ell_a\,\varphi \mid \mathsf{Y}^\ell_a\,\varphi \mid \mathsf{G}^\ell_{\bar a} \mid \mathsf{H}^\ell_{\bar a}$$

for $a \in \Gamma$ and $\varphi, \psi$ are formulas in unambiguous TL. The atomic formulas are $\top$ (which is *true*), and the eager modalities $\mathsf{G}_{\bar a}$ (for Globally-no-$a$) and $\mathsf{H}_{\bar a}$ (for Historically-no-$a$), as well as the lazy modalities $\mathsf{G}^\ell_{\bar a}$ (for lazy-Globally-no-$a$) and $\mathsf{H}^\ell_{\bar a}$ (for lazy-Historically-no-$a$). We now define, when a word $\alpha$ at a position $x \in \mathbb{N} \cup \{\infty\}$ satisfies a formula $\varphi$ in unambiguous TL, in which case we write $\alpha, x \models \varphi$. The atomic formula $\top$ is true at all positions and the semantics of the Boolean connectives is as usual. For $\mathsf{Z} \in \{\mathsf{X}_a, \mathsf{Y}_a, \mathsf{X}^\ell_a, \mathsf{Y}^\ell_a \mid a \in \Gamma\}$ the semantics is defined as follows:

$$\alpha, x \models \mathsf{Z}\,\varphi \quad \text{iff} \quad \mathsf{Z}(x) \text{ is defined and } \alpha, \mathsf{Z}(x) \models \varphi,$$

and the semantics of the atomic modalities is given by

$$\begin{aligned} \mathsf{G}_{\bar a} &= \neg\, \mathsf{X}_a \top, & \mathsf{H}_{\bar a} &= \neg\, \mathsf{Y}_a \top, \\ \mathsf{G}^{\ell}_{\bar a} &= \neg\, \mathsf{X}^{\ell}_a \top, & \mathsf{H}^{\ell}_{\bar a} &= \neg\, \mathsf{Y}^{\ell}_a \top. \end{aligned}$$

In order to define when a word $\alpha$ models a formula $\varphi$, we have to distinguish whether $\varphi$ starts with a future or a past modality:

$$\begin{array}{llllll} \alpha \models \mathsf{X}_a \varphi & \text{iff} & \alpha, 0 \models \mathsf{X}_a \varphi, & \alpha \models \mathsf{Y}_a \varphi & \text{iff} & \alpha, \infty \models \mathsf{Y}_a \varphi, \\ \alpha \models \mathsf{G}_{\bar a} & \text{iff} & \alpha, 0 \models \mathsf{G}_{\bar a}, & \alpha \models \mathsf{H}_{\bar a} & \text{iff} & \alpha, \infty \models \mathsf{H}_{\bar a}, \\ \alpha \models \mathsf{X}^{\ell}_a \varphi & \text{iff} & \alpha, 0 \models \mathsf{X}^{\ell}_a \varphi, & \alpha \models \mathsf{Y}^{\ell}_a \varphi & \text{iff} & \alpha, \infty \models \mathsf{Y}^{\ell}_a \varphi, \\ \alpha \models \mathsf{G}^{\ell}_{\bar a} & \text{iff} & \alpha, 0 \models \mathsf{G}^{\ell}_{\bar a}, & \alpha \models \mathsf{H}^{\ell}_{\bar a} & \text{iff} & \alpha, \infty \models \mathsf{H}^{\ell}_{\bar a}. \end{array}$$

The modalities on the left are called *future modalities* and the modalities on the right are called *past modalities*. The atomic modalities $\mathsf{G}_{\bar a}$ and $\mathsf{G}^{\ell}_{\bar a}$ differ only for the infinite position, but the semantics of $\mathsf{H}_{\bar a}$ and $\mathsf{H}^{\ell}_{\bar a}$ differs a lot: $\alpha \models \mathsf{H}_{\bar a}$ if and only if $a \in \mathrm{im}(\alpha)$ or $a \notin \mathrm{alph}(\alpha)$ whereas $\alpha \models \mathsf{H}^{\ell}_{\bar a}$ if and only if $a \notin \mathrm{alph}(\alpha)$. Every formula $\varphi$ defines a language $L(\varphi) = \{\alpha \in \Gamma^\infty \mid \alpha \models \varphi\}$.

For $\mathcal{C} \subseteq \{\mathsf{X}_a, \mathsf{Y}_a, \mathsf{G}_{\bar a}, \mathsf{H}_{\bar a}, \mathsf{X}^{\ell}_a, \mathsf{Y}^{\ell}_a, \mathsf{G}^{\ell}_{\bar a}, \mathsf{H}^{\ell}_{\bar a}\}$ we define the following fragments of unambiguous TL:

- $\mathrm{TL}[\mathcal{C}]$ consists of all formulas using only $\top$, Boolean connectives, and temporal modalities in $\mathcal{C}$,
- $\mathrm{TL}^+[\mathcal{C}]$ consists of all formulas using only $\top$, positive Boolean connectives (i.e., no negation), and temporal modalities in $\mathcal{C}$,
- $\mathrm{TL}_{\mathsf{X}}[\mathcal{C}]$ consists of all formulas using only $\top$, Boolean connectives, and temporal modalities in $\mathcal{C}$ such that all outmost modalities are future modalities,
- $\mathrm{TL}^+_{\mathsf{X}}[\mathcal{C}]$ consists of all formulas in $\mathrm{TL}^+[\mathcal{C}] \cap \mathrm{TL}_{\mathsf{X}}[\mathcal{C}]$.

Frequently, we identify a class of formulas $\mathcal{F}$ with the class of languages $\{L(\varphi) \mid \varphi \in \mathcal{F}\}$. We say that a language $L \subseteq \Gamma^\infty$ is definable in a logical fragment $\mathcal{F}$ or simply $\mathcal{F}$-definable, if $L = L(\varphi)$ for some $\varphi \in \mathcal{F}$.

**Example 6** Consider again the language $L \subseteq \Gamma^\infty$ of example 5 consisting of all non-empty words with $a$ as the first letter. This language is defined by each of following formulas:

$$\begin{aligned} \varphi_1 &= \mathsf{X}_a \top \wedge \bigwedge_{b \in \Gamma} \neg\, \mathsf{X}_a \mathsf{Y}_b \top && \in \mathrm{TL}_{\mathsf{X}}[\mathsf{X}_a, \mathsf{Y}_a], \\ \varphi_2 &= \bigwedge_{b \in \Gamma} \mathsf{X}_a \mathsf{H}_{\bar b} && \in \mathrm{TL}^+_{\mathsf{X}}[\mathsf{X}_a, \mathsf{H}_{\bar a}], \\ \varphi_3 &= \mathsf{X}_a \top \wedge \bigwedge_{b \in \Gamma \setminus \{a\}} \left( \mathsf{G}_{\bar b} \vee \mathsf{X}_b \mathsf{Y}_a \top \right) && \in \mathrm{TL}^+_{\mathsf{X}}[\mathsf{X}_a, \mathsf{Y}_a, \mathsf{G}_{\bar a}]. \end{aligned}$$

The formula $\varphi_1$ says that there is some $a$-position and that no letter occurs before the first $a$-position. In particular, it uses a negation. The second formula $\varphi_2$ is almost identical, but it uses the atomic modality $\mathsf{H}_{\bar b}$. Due to the use of this implicit negation in a past-modality, no explicit negation is required. The surprising fact about $\varphi_3$ is that it neither uses negations nor the implicitly negated past-modality $\mathsf{H}_{\bar b}$. It essentially says that before every non-$a$-position there is an $a$-position. ◊

**Example 7** The language $L = (\Gamma \setminus \{b\})^* a \Gamma^\infty$ with $a \neq b$ consisting of all words containing an $a$-position with no $b$ to the left is defined by each of the following formulas:

$$\begin{aligned} \varphi_1 &= \mathsf{X}_a \neg \mathsf{Y}_b \top && \in \mathrm{TL}[\mathsf{X}_a, \mathsf{Y}_a], \\ \varphi_2 &= \mathsf{G}_{\bar{b}} \vee \mathsf{X}_b \mathsf{Y}_a \top && \in \mathrm{TL}^+[\mathsf{X}_a, \mathsf{Y}_a, \mathsf{G}_{\bar{a}}]. \end{aligned}$$

The first formula requires that the first $a$-position has no $b$-position in the future, whereas the second formula states that there is either no $b$ at all or that there is an $a$-position before the first $b$-position. Note that for a word in $L$, the position reached by the term $\mathsf{X}_b \mathsf{Y}_a$ in $\varphi_2$ is not necessarily the first $a$-position of the word. In particular, formulas can be equivalent without visiting the same positions. Also note that the argumentation would not be valid for $a = b$. $\Diamond$

Inspired by the atomic logical modalities, we extend the notion of a ranker by allowing the atomic modalities $\mathsf{G}_{\bar{a}}$ and $\mathsf{H}_{\bar{a}}$ as well as $\mathsf{G}^\ell_{\bar{a}}$ and $\mathsf{H}^\ell_{\bar{a}}$. We call $r$ a *ranker with atomic modality* $\mathsf{G}_{\bar{a}}$ ($\mathsf{H}_{\bar{a}}$, $\mathsf{G}^\ell_{\bar{a}}$, $\mathsf{H}^\ell_{\bar{a}}$, resp.) if $r = s\,\mathsf{G}_{\bar{a}}$ ($r = s\,\mathsf{H}_{\bar{a}}$, $r = s\,\mathsf{G}^\ell_{\bar{a}}$, $r = s\,\mathsf{H}^\ell_{\bar{a}}$, resp.) for some ranker $s$. In this setting, $r = \mathsf{G}_{\bar{a}}$ is an $\mathsf{X}$-ranker, and $r = \mathsf{H}_{\bar{a}}$ is a $\mathsf{Y}$-ranker. Analogously, we can add atomic modalities to lazy rankers. Note that any ranker with some atomic modality is also a formula in unambiguous TL. We can therefore define the domain of an extended ranker $r$ with some atomic modality by

$$r(\alpha, x) \text{ is defined} \quad \text{iff} \quad \alpha, x \models r.$$

If $r \in s\,\{\mathsf{G}_{\bar{a}}, \mathsf{H}_{\bar{a}}, \mathsf{G}^\ell_{\bar{a}}, \mathsf{H}^\ell_{\bar{a}} \mid a \in \Gamma\}$ is an extended ranker and $r(\alpha, x)$ is defined, then we set $r(\alpha, x) = s(\alpha, x)$, i.e., $r(\alpha, x)$ is the position reached after the execution of $s$. The reinterpretation of rankers as formulas also makes sense for a ranker $r \in \{\mathsf{X}_a, \mathsf{Y}_a, \mathsf{X}^\ell_a, \mathsf{Y}^\ell_a\}^*$ without atomic modality, if we identify $r$ with $r\top$ in unambiguous TL, which is justified since we have that $r$ is defined on $\alpha$ if and only if $\alpha \models r\top$.

Let $\mathcal{C} \subseteq \{\mathsf{G}_{\bar{a}}, \mathsf{H}_{\bar{a}}, \mathsf{G}^\ell_{\bar{a}}, \mathsf{H}^\ell_{\bar{a}}\}$. A language $L$ is a *ranker language with atomic modalities* $\mathcal{C}$ if $L$ is a Boolean combination of languages $L(r)$ such that $r$ is either a ranker without atomic modalities or a ranker with some atomic modality in $\mathcal{C}$. Similarly, the notions of lazy / positive / $\mathsf{X}$-ranker languages are adapted to the use of atomic modalities.

The following lemma shows that not only can we interpret rankers as formulas, but we can also transform fragments of unambiguous TL into ranker languages.

**Lemma 1** *For $L \subseteq \Gamma^\infty$ the following holds:*

1. *If $L \in \mathrm{TL}[\mathsf{X}_a, \mathsf{Y}_a]$, then $L$ is a ranker language.*
2. *If $L \in \mathrm{TL}^+[\mathsf{X}_a, \mathsf{Y}_a, \mathsf{G}_{\bar{a}}, \mathsf{H}_{\bar{a}}]$, then $L$ is a positive ranker language with atomic modalities $\mathsf{G}_{\bar{a}}$ and $\mathsf{H}_{\bar{a}}$.*
3. *If $L \in \mathrm{TL}^+[\mathsf{X}_a, \mathsf{Y}_a, \mathsf{G}_{\bar{a}}]$, then $L$ is a positive ranker language with atomic modality $\mathsf{G}_{\bar{a}}$.*
4. *If $L \in \mathrm{TL}^+_{\mathsf{X}}[\mathsf{X}_a, \mathsf{Y}_a, \mathsf{G}_{\bar{a}}]$, then $L$ is a positive $\mathsf{X}$-ranker language with atomic modality $\mathsf{G}_{\bar{a}}$.*
5. *If $L \in \mathrm{TL}_{\mathsf{X}}[\mathsf{X}_a, \mathsf{Y}_a]$, then $L$ is an $\mathsf{X}$-ranker language.*
6. *If $L \in \mathrm{TL}^+[\mathsf{X}^\ell_a, \mathsf{Y}^\ell_a, \mathsf{H}^\ell_{\bar{a}}]$, then $L$ is a positive lazy ranker language with atomic modality $\mathsf{H}_{\bar{a}}$.*

*Proof:* We observe the following basic equivalences (with $\mathsf{Z}_a \in \{\mathsf{X}_a, \mathsf{Y}_a, \mathsf{X}_a^\ell, \mathsf{Y}_a^\ell\}$ and with $\equiv$ denoting equivalence of formulas on all words and all positions in $\mathbb{N} \cup \{\infty\}$):

$$\begin{aligned} \mathsf{Z}_a(\neg\varphi) &\equiv \mathsf{Z}_a \top \wedge \neg \mathsf{Z}_a \varphi, \\ \mathsf{Z}_a(\varphi \vee \psi) &\equiv \mathsf{Z}_a \varphi \vee \mathsf{Z}_a \psi, \\ \mathsf{Z}_a(\varphi \wedge \psi) &\equiv \mathsf{Z}_a \varphi \wedge \mathsf{Z}_a \psi. \end{aligned}$$

For a formula in $\mathrm{TL}[\mathsf{X}_a, \mathsf{Y}_a]$ we use the equivalences to move all Boolean connectives to the outermost level, ending up in a Boolean combination of formulas of type $r\top$ for some ranker $r$. This shows 1. For formulas in $\mathrm{TL}^+[\mathsf{X}_a, \mathsf{Y}_a, \mathsf{G}_{\bar{a}}, \mathsf{H}_{\bar{a}}]$ the same argument yields a positive Boolean combination of languages defined by rankers with atomic modalities $\mathsf{G}_{\bar{a}}$ and $\mathsf{H}_{\bar{a}}$. (Of course, we do not apply the rule for negations.) Moreover, there is a ranker generated this way containing the atomic modality $\mathsf{H}_{\bar{a}}$ if and only if the original formula uses $\mathsf{H}_{\bar{a}}$. This shows 2 and 3. The situation for formulas in $\mathrm{TL}^+[\mathsf{X}_a^\ell, \mathsf{Y}_a^\ell, \mathsf{H}_{\bar{a}}^\ell]$ is similar, showing 6. If the first non-Boolean modality on each path of the syntax tree of the original formula is a future modality, then all rankers generated by the above rules start with future modalities, so 4 and 5 follow. □

**Lemma 2** *For every non-empty ranker $r$ there exist formulas $\varrho_r, \vartheta_r \in \mathrm{TL}^+[\mathsf{X}_a, \mathsf{Y}_a, \mathsf{G}_{\bar{a}}]$ such that for every $\alpha \in \Gamma^\infty$ with $r(\alpha)$ being defined we have*

$$\begin{aligned} \alpha, x \models \varrho_r \quad &\textit{iff} \quad x > r(\alpha), \\ \alpha, x \models \vartheta_r \quad &\textit{iff} \quad x \geq r(\alpha). \end{aligned}$$

*Proof:* We use induction on the length of the ranker. Let $\alpha$ be a word such that $r$ is defined on $\alpha$. Consider the case $r = s\,\mathsf{X}_a$ for some ranker $s$. Note that there must be an $a$-position since $r$ is defined on $\alpha$. For a position $x$ of $\alpha$ we have $x > r(\alpha)$ if and only if we find an $a$-position $y$ strictly smaller than $x$ such that $y > s(\alpha)$. But this is equivalent to $\mathsf{Y}_a(x) > s(\alpha)$ since $\mathsf{Y}_a(x)$ is the maximal $a$-position strictly smaller than $x$. For the formula $\vartheta_r$ we have to be more careful: For a position $x$ of $\alpha$ we have $x \geq r(\alpha)$ if and only if there is no $a$-position strictly greater than $x$ or if all such $a$-positions $y$ satisfy $y > r(\alpha)$ which we already know how to express. Therefore,

$$\begin{aligned} \varrho_r &= \mathsf{Y}_a\, \varrho_s \\ \vartheta_r &= \mathsf{G}_{\bar{a}} \vee \mathsf{X}_a\, \varrho_r \end{aligned}$$

for $r = s\,\mathsf{X}_a$ with $\varrho_s = \top$ for $s = \varepsilon$. So, if the ranker starts with $\mathsf{X}_a$, we view the empty ranker to be on a position in front of the word and hence all positions are strictly greater than it.

Consider $r = s\,\mathsf{Y}_a$ for some ranker $s$. Again, we know that there is an $a$-position in $\alpha$. For a position $x$ on $\alpha$ we have $x \geq r(\alpha)$ if and only if there is no $a$-position strictly greater than $x$ or for all such $a$-positions $y$ we have $y \geq s(\alpha)$. But this is equivalent to $\mathsf{X}_a(x) \geq s(\alpha)$ since $\mathsf{X}_a(x)$ is the minimal $a$-position strictly greater than $x$. A position $x$ of $\alpha$ satisfies $x > r(\alpha)$ if and only if there is an $a$-position $y$ strictly smaller than $x$ such that $y \geq r(\alpha)$. Hence

$$\begin{aligned} \vartheta_r &= \mathsf{G}_{\bar{a}} \vee \mathsf{X}_a\, \vartheta_s \\ \varrho_r &= \mathsf{Y}_a\, \vartheta_r \end{aligned}$$

for $r = s\,\mathsf{Y}_a$ with $\vartheta_s = \bot$ for $s = \varepsilon$. This means that in the case that the ranker starts with $\mathsf{Y}_a$, we view the empty ranker to be on a position behind the word and hence all positions are strictly smaller than it. So for $s = \varepsilon$ the formula $\vartheta_r$ is equivalent to $\mathsf{G}_{\bar{a}}$. □

In the following lemma, we set $\infty \leq \infty$ and also $\infty < \infty$. This is natural since with the single "position" $\infty$ we want to model the behavior of a word "after" all finite positions; in particular, if $\mathrm{im}(\alpha) \neq \emptyset$, then $\infty$ corresponds to infinitely many positions.

**Lemma 3** *For every non-empty lazy ranker $r$ there exist formulas $\varrho_r, \vartheta_r \in \mathrm{TL}^+[\mathsf{X}_a^\ell, \mathsf{Y}_a^\ell, \mathsf{H}_{\bar{a}}^\ell]$ such that for every $\alpha \in \Gamma^\infty$ with $r(\alpha)$ being defined we have*

$$\alpha, x \models \varrho_r \quad \text{iff} \quad x < r(\alpha),$$
$$\alpha, x \models \vartheta_r \quad \text{iff} \quad x \leq r(\alpha).$$

*Proof:* For $r = \mathsf{X}_a^\ell$ we set

$$\vartheta_r = \mathsf{H}_{\bar{a}}^\ell\,,$$
$$\varrho_r = \mathsf{X}_a^\ell\, \vartheta_r\,.$$

For $r = \mathsf{Y}_a^\ell$ we set

$$\varrho_r = \mathsf{X}_a^\ell\, \top,$$
$$\vartheta_r = \mathsf{H}_{\bar{a}}^\ell \vee \mathsf{Y}_a^\ell\, \varrho_r\,.$$

Suppose $r = s\, \mathsf{X}_a^\ell$. Then

$$\vartheta_r = \mathsf{H}_{\bar{a}}^\ell \vee \mathsf{Y}_a^\ell\, \vartheta_s\,,$$
$$\varrho_r = \mathsf{X}_a^\ell\, \vartheta_r\,.$$

Suppose $r = s\, \mathsf{Y}_a^\ell$. Then

$$\varrho_r = \mathsf{X}_a^\ell\, \varrho_s$$
$$\vartheta_r = \mathsf{H}_{\bar{a}}^\ell \vee \mathsf{Y}_a^\ell\, \varrho_r\,.$$

Note that the formulas conform to $\infty < \infty$ and $\infty \leq \infty$. □

## 3.3 Unambiguous Interval Temporal Logic

Here, we extend unambiguous interval temporal logic (unambiguous ITL) to infinite words in such a way that it coincides with $\mathrm{FO}^2$. In fact, we have two extensions with this property, one being eager and one being lazy. The syntax of unambiguous ITL is given by Boolean combinations and:

$$\top \mid \varphi\, \mathsf{F}_a\, \psi \mid \varphi\, \mathsf{L}_a\, \psi \mid \mathsf{G}_{\bar{a}} \mid \mathsf{H}_{\bar{a}} \mid \varphi\, \mathsf{F}_a^\ell\, \psi \mid \varphi\, \mathsf{L}_a^\ell\, \psi \mid \mathsf{G}_{\bar{a}}^\ell \mid \mathsf{H}_{\bar{a}}^\ell$$

with $a \in \Gamma$ and $\varphi, \psi$ are formulas in unambiguous ITL. The name $\mathsf{F}_a$ derives from First-$a$ and $\mathsf{L}_a$ derives from Last-$a$. As in unambiguous temporal logic, the atomic formulas are $\top$, the eager modalities $\mathsf{G}_{\bar{a}}$ and $\mathsf{H}_{\bar{a}}$, and the lazy modalities $\mathsf{G}_{\bar{a}}^\ell$ and $\mathsf{H}_{\bar{a}}^\ell$. We now define, when a word $\alpha$ together with an interval $(x; y) = \{z \in \mathbb{N} \cup \{\infty\} \mid x < z < y\}$ satisfies a formula $\varphi$ in unambiguous ITL, in which case we write $\alpha, (x; y) \models \varphi$. Remember that we have set $\infty < \infty$. In particular $(\infty; \infty) = \{\infty\}$. The atomic formula $\top$ is true for all intervals and the semantics of the Boolean connectives is as usual. The semantics of the binary modalities is as follows:

$$
\begin{array}{lll}
\alpha,(x;y) \models \varphi \,\mathsf{F}_a\, \psi & \text{iff} & \mathsf{X}_a(x) \text{ is defined, } \mathsf{X}_a(x) < y, \\
 & & \alpha,\big(x;\mathsf{X}_a(x)\big) \models \varphi \;\text{ and }\; \alpha,\big(\mathsf{X}_a(x);y\big) \models \psi, \\
\alpha,(x;y) \models \varphi \,\mathsf{L}_a\, \psi & \text{iff} & \mathsf{Y}_a(y) \text{ is defined, } \mathsf{Y}_a(y) > x, \\
 & & \alpha,\big(x;\mathsf{Y}_a(y)\big) \models \varphi \;\text{ and }\; \alpha,\big(\mathsf{Y}_a(y);y\big) \models \psi, \\
\alpha,(x;y) \models \varphi \,\mathsf{F}^\ell_a\, \psi & \text{iff} & \mathsf{X}^\ell_a(x) \text{ is defined, } \mathsf{X}^\ell_a(x) < y, \\
 & & \alpha,\big(x;\mathsf{X}^\ell_a(x)\big) \models \varphi \;\text{ and }\; \alpha,\big(\mathsf{X}^\ell_a(x);y\big) \models \psi, \\
\alpha,(x;y) \models \varphi \,\mathsf{L}^\ell_a\, \psi & \text{iff} & \mathsf{Y}^\ell_a(y) \text{ is defined, } \mathsf{Y}^\ell_a(y) > x, \\
 & & \alpha,\big(x;\mathsf{Y}^\ell_a(y)\big) \models \varphi \;\text{ and }\; \alpha,\big(\mathsf{Y}^\ell_a(y);y\big) \models \psi.
\end{array}
$$

The semantics of the atomic modalities is given by

$$
\begin{array}{lllll}
\mathsf{G}_{\bar a} & = \neg(\top \,\mathsf{F}_a\, \top), & \qquad \mathsf{H}_{\bar a} & = \neg(\top \,\mathsf{L}_a\, \top), \\
\mathsf{G}^\ell_{\bar a} & = \neg(\top \,\mathsf{F}^\ell_a\, \top), & \qquad \mathsf{H}^\ell_{\bar a} & = \neg(\top \,\mathsf{L}^\ell_a\, \top) \vee \bigvee_{b\in\Gamma} ((\top \,\mathsf{L}^\ell_b\, \top)\,\mathsf{F}^\ell_b\, \top).
\end{array}
$$

In the definition of $\mathsf{H}^\ell_{\bar a}$, the disjunction on the right-hand side ensures that $\alpha,(\infty;\infty) \models \mathsf{H}^\ell_{\bar a}$ for every infinite word $\alpha \in \Gamma^\omega$ and every $a \in \Gamma$. It will turn out that the inability of specifying the letters not in $\mathrm{im}(\alpha)$ is crucial in the characterization of the fragment $\Pi_2 \cap \mathrm{FO}^2$. Observe that only for the interval $(\infty;\infty)$, there can be a $b$ before the "first" $b$. Also note that for every finite interval of some word $\alpha$, the formula $\mathsf{G}_{\bar a}$ is true if and only if $\mathsf{H}_{\bar a}$ is true. Whether a word $\alpha$ models a formula $\varphi$ in unambiguous ITL (i.e., $\alpha \models \varphi$) or not is defined by

$$\alpha \models \varphi \quad \text{iff} \quad \alpha,(0;\infty) \models \varphi,$$

and the language defined by $\varphi$ is $L(\varphi) = \{\alpha \in \Gamma^\infty \mid \alpha \models \varphi\}$.

Figure 3 depicts the situation for the formula $(\varphi_1 \,\mathsf{F}_b\, \psi_1)\mathsf{L}_a(\varphi_2 \mathsf{F}_c \psi_2)$ being defined on $\alpha$. The main difference to rankers and unambiguous TL is that there is no crossing over in unambiguous ITL, e.g., in the situation depicted on the left side of Figure 2, the formula $(\top \mathsf{L}_b(\top \mathsf{F}_c \top))\mathsf{L}_a\top$ is false even though $\mathsf{Y}_a\,\mathsf{Y}_b\,\mathsf{X}_c$ is defined.

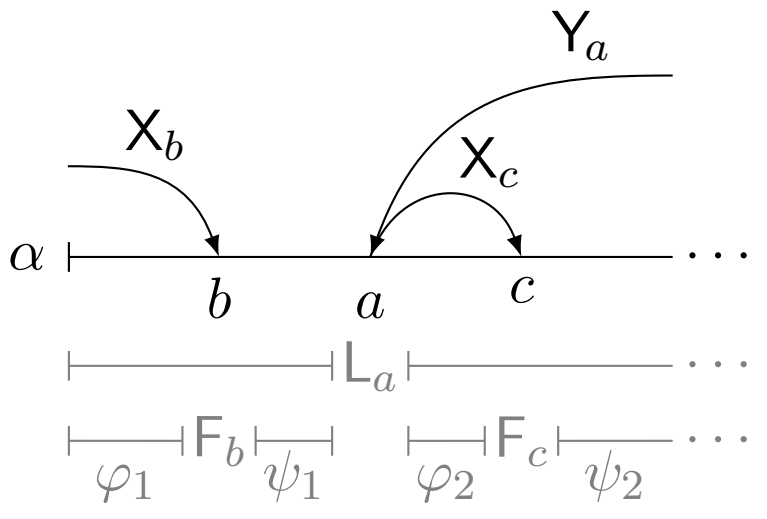

Figure 3: $(\varphi_1 \,\mathsf{F}_b\, \psi_1)\,\mathsf{L}_a\,(\varphi_2 \,\mathsf{F}_c\, \psi_2)$

In unambiguous ITL, the modalities $\mathsf{F}_a, \mathsf{G}_{\bar a}, \mathsf{F}^\ell_a, \mathsf{G}^\ell_{\bar a}$ are *future modalities* and $\mathsf{L}_a, \mathsf{H}_{\bar a}, \mathsf{L}^\ell_a, \mathsf{H}^\ell_{\bar a}$ are *past modalities*. An unambiguous ITL formula $\varphi$ is a *future-formula* if in the parse tree of $\varphi$ every past modality occurs on the left branch of some future modality, i.e., if it is never necessary to interpret a past modality over an unbounded interval. For $\mathcal{C} \subseteq \{\mathsf{F}_a, \mathsf{L}_a, \mathsf{G}_{\bar a}, \mathsf{H}_{\bar a}, \mathsf{F}^\ell_a, \mathsf{L}^\ell_a, \mathsf{G}^\ell_{\bar a}, \mathsf{H}^\ell_{\bar a}\}$ we define the following fragments of unambiguous ITL:

- $\mathrm{ITL}[\mathcal{C}]$ consists of all formulas using only $\top$, Boolean connectives, and temporal modalities in $\mathcal{C}$,
- $\mathrm{ITL}^+[\mathcal{C}]$ consists of all formulas using only $\top$, positive Boolean connectives (i.e., no negation), and temporal modalities in $\mathcal{C}$,
- $\mathrm{ITL}_{\mathsf{F}}[\mathcal{C}]$ consists of all future formulas using only $\top$, Boolean connectives, and temporal modalities in $\mathcal{C}$,
- $\mathrm{ITL}^+_{\mathsf{F}}[\mathcal{C}]$ consists of all future formulas in $\mathrm{ITL}^+[\mathcal{C}] \cap \mathrm{ITL}_{\mathsf{F}}[\mathcal{C}]$.

**Example 8** Consider the unambiguous ITL formulas

$$\varphi = (\top \, \mathsf{L}_b^\ell \, \top) \, \mathsf{F}_b^\ell \, \top \quad \text{and} \quad \psi = \top \, \mathsf{L}_b^\ell \, (\top \, \mathsf{F}_b^\ell \, \top)$$

and a word $\alpha$ with $b \in \mathrm{im}(\alpha)$. The formula $\varphi$ was used in the definition of the semantics of $\mathsf{H}_{\bar{a}}^\ell$ and, as already mentioned, is only true if the interval is $(\infty; \infty)$. In contrast, $\psi$ is also true if the interval is $(0, \infty)$. $\Diamond$

The following two propositions describe a procedure for converting unambiguous ITL formulas into unambiguous TL formulas without introducing new negations. A similar relativization technique as in our proof has been used by Lodaya, Pandya, and Shah [4] for the conversion of ITL over finite words into so-called *deterministic partially ordered two-way automata* (but without the focus on not introducing negations). Proposition 1 is the eager version, whereas Proposition 2 is the lazy version. As will follow from Theorems 1 to 5 we actually have equality for all inclusions in both propositions.

**Proposition 1** *We have the following inclusions:*

$$\begin{aligned}
\mathrm{ITL}[\mathsf{F}_a, \mathsf{L}_a] &\subseteq \mathrm{TL}[\mathsf{X}_a, \mathsf{Y}_a], \\
\mathrm{ITL}^+[\mathsf{F}_a, \mathsf{L}_a, \mathsf{G}_{\bar{a}}, \mathsf{H}_{\bar{a}}] &\subseteq \mathrm{TL}^+[\mathsf{X}_a, \mathsf{Y}_a, \mathsf{G}_{\bar{a}}, \mathsf{H}_{\bar{a}}], \\
\mathrm{ITL}^+[\mathsf{F}_a, \mathsf{L}_a, \mathsf{G}_{\bar{a}}] &\subseteq \mathrm{TL}^+[\mathsf{X}_a, \mathsf{Y}_a, \mathsf{G}_{\bar{a}}], \\
\mathrm{ITL}^+_{\mathsf{F}}[\mathsf{F}_a, \mathsf{L}_a, \mathsf{G}_{\bar{a}}, \mathsf{H}_{\bar{a}}] &\subseteq \mathrm{TL}^+_{\mathsf{X}}[\mathsf{X}_a, \mathsf{Y}_a, \mathsf{G}_{\bar{a}}], \\
\mathrm{ITL}_{\mathsf{F}}[\mathsf{F}_a, \mathsf{L}_a] &\subseteq \mathrm{TL}_{\mathsf{X}}[\mathsf{X}_a, \mathsf{Y}_a].
\end{aligned}$$

*Proof:* Note that the atomic modalities $\mathsf{G}_{\bar{a}}$ and $\mathsf{H}_{\bar{a}}$ are expressible in $\mathrm{ITL}[\mathsf{F}_a, \mathsf{L}_a]$ as well as in $\mathrm{TL}[\mathsf{X}_a, \mathsf{Y}_a]$. For every $\varphi \in \mathrm{ITL}[\mathsf{F}_a, \mathsf{L}_a, \mathsf{G}_{\bar{a}}, \mathsf{H}_{\bar{a}}]$ we construct an equivalent formula $\varphi_{(\varepsilon;\varepsilon)} \in \mathrm{TL}[\mathsf{X}_a, \mathsf{Y}_a, \mathsf{G}_{\bar{a}}, \mathsf{H}_{\bar{a}}]$ such that $\varphi_{(\varepsilon;\varepsilon)}$ contains a negation if and only if $\varphi$ contains a negation, and an $\mathsf{H}_{\bar{a}}$ term appears in $\varphi_{(\varepsilon;\varepsilon)}$ if and only if it appears in $\varphi$. This will prove the first three inclusions.

For rankers $q, r$ and $\varphi \in \mathrm{ITL}[\mathsf{F}_a, \mathsf{L}_a, \mathsf{G}_{\bar{a}}, \mathsf{H}_{\bar{a}}]$ we define $\varphi_{(q;r)} \in \mathrm{TL}[\mathsf{X}_a, \mathsf{Y}_a, \mathsf{G}_{\bar{a}}, \mathsf{H}_{\bar{a}}]$ such that $\alpha \models \varphi_{(q;r)}$ if and only if $q(\alpha)$ and $r(\alpha)$ are defined, $q(\alpha) < r(\alpha)$, and

$$\alpha, \big(q(\alpha); r(\alpha)\big) \models \varphi$$

with $q(\alpha) = 0$ for $q = \varepsilon$ and $r(\alpha) = \infty$ for $r = \varepsilon$. In particular, in the above situation $q$ and $r$ define the boundaries of an interval $(q; r)$ parameterized by words $\alpha$. The construction of $\varphi_{(q;r)}$ is by structural induction. We will make extensive use of the formulas $\varrho_q$, $\varrho_r$ and $\vartheta_r$ from Lemma 2 with the convention $\varrho_q = \top$ for $q = \varepsilon$. The atomic formula $\top$ and Boolean connectives are as follows:

$$\begin{aligned}
\top_{(q;r)} &= q\top \wedge r\top \wedge r\varrho_q \\
(\neg\varphi)_{(q;r)} &= \top_{(q;r)} \wedge \neg\varphi_{(q;r)} \\
(\varphi \wedge \psi)_{(q;r)} &= \varphi_{(q;r)} \wedge \psi_{(q;r)} \\
(\varphi \vee \psi)_{(q;r)} &= \varphi_{(q;r)} \vee \psi_{(q;r)}
\end{aligned}$$

For the atomic ITL-formula $\mathsf{G}_{\bar{a}}$ we set

$$(\mathsf{G}_{\bar{a}})_{(q;r)} = \begin{cases} \top_{(q;\varepsilon)} \wedge q \, \mathsf{G}_{\bar{a}} & \text{for } r = \varepsilon \\ \top_{(q;r)} \wedge q\big(\, \mathsf{G}_{\bar{a}} \vee \mathsf{X}_a \, \vartheta_r \big) & \text{for } r \neq \varepsilon \end{cases}$$

Essentially, the term on the right-hand side says that the next $a$-position after the $q$-position is at least the $r$-position. In case of the atomic ITL-formula $\mathsf{H}_{\bar{a}}$ we have to distinguish between $r = \varepsilon$ and $r \neq \varepsilon$. If $r \neq \varepsilon$, then the interval defined by $(q;r)$ is finite and therefore in this situation $\mathsf{H}_{\bar{a}}$ and $\mathsf{G}_{\bar{a}}$ are equivalent:

$$(\mathsf{H}_{\bar{a}})_{(q;r)} \;=\; (\mathsf{G}_{\bar{a}})_{(q;r)}$$

If $r = \varepsilon$, then

$$(\mathsf{H}_{\bar{a}})_{(q;r)} \;=\; \top_{(q;r)} \wedge \big(\,\mathsf{H}_{\bar{a}} \vee q\,\mathsf{G}_{\bar{a}}\,\big)$$

i.e., either there are no or infinitely many $a$-positions, or there is no $a$-position after the $q$-position. For the $\mathsf{F}_a$-modality we define

$$(\varphi\,\mathsf{F}_a\,\psi)_{(q;r)} \;=\; \top_{(q;r)} \wedge \varphi_{(q;q\,\mathsf{X}_a)} \wedge \psi_{(q\,\mathsf{X}_a;r)}$$

i.e., we verify $\varphi$ on the interval $(q;q\,\mathsf{X}_a)$ and $\psi$ on the interval $(q\,\mathsf{X}_a;r)$. The $\mathsf{L}_a$-modality is similar:

$$(\varphi\,\mathsf{L}_a\,\psi)_{(q;r)} \;=\; \top_{(q;r)} \wedge \varphi_{(q;r\,\mathsf{Y}_a)} \wedge \psi_{(r\,\mathsf{Y}_a;r)}$$

saying that $\varphi$ and $\psi$ are defined on the respective subintervals and that there is some $a$-position in the interval $(q;r)$.

Now, for every $\varphi \in \mathrm{ITL}[\mathsf{F}_a, \mathsf{L}_a, \mathsf{G}_{\bar{a}}, \mathsf{H}_{\bar{a}}]$ and any rankers $q, r$, the formula $\varphi_{(q;r)} \in \mathrm{TL}[\mathsf{X}_a, \mathsf{Y}_a, \mathsf{G}_{\bar{a}}, \mathsf{H}_{\bar{a}}]$ is a Boolean combination of formulas of the form $\top_{(p;s)}$, $(\mathsf{G}_{\bar{a}})_{(p;s)}$, and $(\mathsf{H}_{\bar{a}})_{(p;s)}$. Moreover, every negation and every $\mathsf{H}_{\bar{a}}$-modality in $\varphi_{(q;r)}$ is only caused by the respective operation in $\varphi$. This completes the proof of the first three inclusions.

For the last two inclusions, we first observe that in our construction the following invariants hold:

- If $q$ and $r$ are $\mathsf{X}$-rankers with $r \neq \varepsilon$, then $\varphi_{(q;r)} \in \mathrm{TL}_{\mathsf{X}}[\mathsf{X}_a, \mathsf{Y}_a, \mathsf{G}_{\bar{a}}]$ for every formula $\varphi \in \mathrm{ITL}[\mathsf{F}_a, \mathsf{L}_a, \mathsf{G}_{\bar{a}}, \mathsf{H}_{\bar{a}}]$.
- If $q$ is an $\mathsf{X}$-ranker and $r = \varepsilon$, then for every $\varphi \in \mathrm{ITL}[\mathsf{F}_a, \mathsf{L}_a, \mathsf{G}_{\bar{a}}, \mathsf{H}_{\bar{a}}]$ and every $\psi \in \mathrm{ITL}_{\mathsf{F}}[\mathsf{F}_a, \mathsf{L}_a, \mathsf{G}_{\bar{a}}, \mathsf{H}_{\bar{a}}]$ we have $(\varphi\,\mathsf{F}_a\,\psi)_{(q;r)} \in \mathrm{TL}_{\mathsf{X}}[\mathsf{X}_a, \mathsf{Y}_a, \mathsf{G}_{\bar{a}}]$.

Therefore, if $\varphi \in \mathrm{ITL}_{\mathsf{F}}[\mathsf{F}_a, \mathsf{L}_a, \mathsf{G}_{\bar{a}}, \mathsf{H}_{\bar{a}}]$, then $\varphi_{(\varepsilon;\varepsilon)} \in \mathrm{TL}_{\mathsf{X}}[\mathsf{X}_a, \mathsf{Y}_a, \mathsf{G}_{\bar{a}}]$. Hence, the fourth and the fifth inclusion follow. □

**Proposition 2** *We have the following inclusions:*

$$\begin{aligned}
\mathrm{ITL}[\mathsf{F}_a^\ell, \mathsf{L}_a^\ell] &\subseteq \mathrm{TL}[\mathsf{X}_a^\ell, \mathsf{Y}_a^\ell], \\
\mathrm{ITL}^+[\mathsf{F}_a^\ell, \mathsf{L}_a, \mathsf{G}_{\bar{a}}^\ell, \mathsf{H}_{\bar{a}}^\ell] &\subseteq \mathrm{TL}^+[\mathsf{X}_a^\ell, \mathsf{Y}_a^\ell, \mathsf{G}_{\bar{a}}^\ell, \mathsf{H}_{\bar{a}}^\ell], \\
\mathrm{ITL}^+[\mathsf{F}_a^\ell, \mathsf{L}_a^\ell, \mathsf{H}_{\bar{a}}^\ell] &\subseteq \mathrm{TL}^+[\mathsf{X}_a^\ell, \mathsf{Y}_a^\ell, \mathsf{H}_{\bar{a}}^\ell].
\end{aligned}$$

*Proof:* For $\varphi \in \mathrm{ITL}[\mathsf{F}_a^\ell, \mathsf{L}_a^\ell, \mathsf{G}_{\bar{a}}^\ell, \mathsf{H}_{\bar{a}}^\ell]$ we construct an equivalent formula $\varphi_{(\varepsilon;\varepsilon)} \in \mathrm{TL}[\mathsf{X}_a^\ell, \mathsf{Y}_a^\ell, \mathsf{G}_{\bar{a}}^\ell, \mathsf{H}_{\bar{a}}^\ell]$ such that $\varphi_{(\varepsilon;\varepsilon)}$ contains a negation if and only if $\varphi$ does, and a $\mathsf{G}_{\bar{a}}^\ell$ term appears in $\varphi_{(\varepsilon;\varepsilon)}$ if and only if it appears in $\varphi$. We use the following construction. For lazy rankers $q, r$ and $\varphi \in \mathrm{ITL}[\mathsf{F}_a^\ell, \mathsf{L}_a^\ell, \mathsf{G}_{\bar{a}}^\ell, \mathsf{H}_{\bar{a}}^\ell]$ we define $\varphi_{(q;r)} \in \mathrm{TL}[\mathsf{X}_a^\ell, \mathsf{Y}_a^\ell, \mathsf{G}_{\bar{a}}^\ell, \mathsf{H}_{\bar{a}}^\ell]$ such that $\alpha \models \varphi_{(q;r)}$ if and only if $q(\alpha)$ and $r(\alpha)$ are defined, $q(\alpha) < r(\alpha)$, and

$$\alpha, \big(q(\alpha); r(\alpha)\big) \models \varphi$$

with $q(\alpha) = 0$ for $q = \varepsilon$ and $r(\alpha) = \infty$ for $r = \varepsilon$. In particular, in the above situation $q$ and $r$ define the boundaries of an interval $(q;r)$ parameterized by words $\alpha$. The construction of $\varphi_{(q;r)}$ is by structural induction. We will make extensive use of the formulas $\varrho_r$ and $\vartheta_r$ from Lemma 3. The atomic formula $\top$ and positive Boolean connectives are as follows:

$$\begin{aligned} \top_{(q;r)} &= q\top \wedge r\top \wedge q\varrho_r \\ (\neg\varphi)_{(q;r)} &= \top_{(q;r)} \wedge \neg\varphi_{(q;r)} \\ (\varphi \wedge \psi)_{(q;r)} &= \varphi_{(q;r)} \wedge \psi_{(q;r)} \\ (\varphi \vee \psi)_{(q;r)} &= \varphi_{(q;r)} \vee \psi_{(q;r)} \end{aligned}$$

with $\varrho_r = \top$ for $r = \varepsilon$. For the atomic ITL-formula $\mathsf{H}^\ell_{\bar{a}}$ we set

$$(\mathsf{H}^\ell_{\bar{a}})_{(q;r)} = \begin{cases} \top_{(q;r)} \wedge r(\mathsf{H}^\ell_{\bar{a}} \vee \mathsf{Y}^\ell_a \vartheta_q) & \text{if } q \neq \varepsilon \\ \top_{(\varepsilon;r)} \wedge r\,\mathsf{H}^\ell_{\bar{a}} & \text{if } q = \varepsilon \end{cases}$$

This is consistent with the definition that for all $a \in \Gamma$ and all infinite words $\alpha \in \Gamma^\omega$ we have $\alpha, (\infty;\infty) \models \mathsf{H}^\ell_{\bar{a}}$ in unambiguous ITL. The atomic modality $\mathsf{G}_{\bar{a}}$ is slightly more technical due to its behavior on the interval $(\infty;\infty)$. For $r = \varepsilon$ we set

$$(\mathsf{G}^\ell_{\bar{a}})_{(q;\varepsilon)} = \top_{(q;\varepsilon)} \wedge q\,\mathsf{G}^\ell_{\bar{a}}\ .$$

If $r \neq \varepsilon$ is an $\mathsf{X}^\ell$-ranker we define

$$(\mathsf{G}^\ell_{\bar{a}})_{(q;r)} = \top_{(q;r)} \wedge r(\mathsf{H}^\ell_{\bar{a}} \vee \mathsf{Y}^\ell_a \vartheta_q)$$

with $\vartheta_q = \bot$ for $q = \varepsilon$. Therefore, if $q = \varepsilon$, then we can omit the term $\mathsf{Y}^\ell_a \vartheta_q$ in the above formula. Finally, for a non-empty $\mathsf{Y}^\ell$-ranker $r$ we use

$$(\mathsf{G}^\ell_{\bar{a}})_{(q;r)} = \top_{(q;r)} \wedge \left( \left( r(\mathsf{H}^\ell_{\bar{a}} \vee \mathsf{Y}^\ell_a \vartheta_q) \wedge \bigvee_{b \in B} \mathsf{Y}^\ell_b \mathsf{G}^\ell_{\bar{b}} \right) \vee q\,\mathsf{G}^\ell_{\bar{a}} \right)$$

with $B = \mathrm{alph}_\Gamma(r)$ is the set of letters which occur in some modality of the ranker $r$. As before, we set $\vartheta_q = \bot$ for $q = \varepsilon$. The above formula distinguishes two cases. The first case is that $r$ determines a finite position (after some last occurrence of a letter $b \in B$ from the ranker $r$ there is no $b$-position). In this case either there is no $a$-position before the $r$-position, or the first $a$-position before the $r$-position is on the left-hand side of the $q$-position. The other case is that $r$ leads to the infinite position and then we want to see no $a$-position on the right-hand side of the $q$-position.

For the $\mathsf{F}^\ell_a$-modality we define

$$(\varphi\,\mathsf{F}^\ell_a\,\psi)_{(q;r)} = \top_{(q;r)} \wedge \varphi_{(q;q\,\mathsf{X}^\ell_a)} \wedge \psi_{(q\,\mathsf{X}^\ell_a;r)}$$

and for $\mathsf{L}^\ell_a$ we set

$$(\varphi\,\mathsf{L}^\ell_a\,\psi)_{(q;r)} = \top_{(q;r)} \wedge \varphi_{(q;r\,\mathsf{Y}^\ell_a)} \wedge \psi_{(r\,\mathsf{Y}^\ell_a;r)}\,.$$

Now, for every $\varphi \in \mathrm{ITL}[\mathsf{F}^\ell_a, \mathsf{L}^\ell_a, \mathsf{G}^\ell_{\bar{a}}, \mathsf{H}^\ell_{\bar{a}}]$ and for all lazy rankers $q, r$, the formula $\varphi_{(q;r)} \in \mathrm{TL}[\mathsf{X}^\ell_a, \mathsf{Y}^\ell_a, \mathsf{G}^\ell_{\bar{a}}, \mathsf{H}^\ell_{\bar{a}}]$ is a Boolean combination of formulas of the form $\top_{(p;s)}$, $(\mathsf{G}^\ell_{\bar{a}})_{(p;s)}$, and $(\mathsf{H}^\ell_{\bar{a}})_{(p;s)}$. Moreover, every negation and every $\mathsf{G}^\ell_{\bar{a}}$-modality in $\varphi_{(q;r)}$ is due to the respective operation in $\varphi$. This completes the proof. □

# 4 The Fragment $\mathrm{FO}^2$

This section contains various ITL, TL, and ranker characterizations using the eager variants. We postpone characterizations in terms of the lazy fragments to Theorem 5.

**Theorem 1** *For $L \subseteq \Gamma^\infty$ the following assertions are equivalent:*

1. *$L$ is definable in* $\mathrm{FO}^2$.
2. *$L$ is definable in* $\mathrm{ITL}^+[\mathsf{F}_a, \mathsf{L}_a, \mathsf{G}_{\bar{a}}, \mathsf{H}_{\bar{a}}]$.
3. *$L$ is definable in* $\mathrm{ITL}[\mathsf{F}_a, \mathsf{L}_a]$.
4. *$L$ is definable in* $\mathrm{TL}[\mathsf{X}_a, \mathsf{Y}_a]$.
5. *$L$ is definable in* $\mathrm{TL}^+[\mathsf{X}_a, \mathsf{Y}_a, \mathsf{G}_{\bar{a}}, \mathsf{H}_{\bar{a}}]$.
6. *$L$ is a positive ranker language with atomic modalities $\mathsf{G}_{\bar{a}}$ and $\mathsf{H}_{\bar{a}}$.*
7. *$L$ is a ranker language.*

**Lemma 4** *Let $A \subseteq \Gamma$. Then $A^\infty$ is definable in $\mathrm{ITL}^+[\mathsf{G}_{\bar{a}}]$, and $A^{\mathrm{im}}$ is definable in $\mathrm{ITL}^+[\mathsf{F}_a, \mathsf{H}_{\bar{a}}]$.*

*Proof:* A letter $a \in \Gamma$ does not appear in a word $\alpha$ if and only if $\alpha \models \mathsf{G}_{\bar{a}}$, and $a$ appears infinitely often in a word $\alpha$ if and only if $\alpha \models (\top \mathsf{F}_a \top) \wedge \mathsf{H}_{\bar{a}}$. Hence

$$A^\infty \text{ is defined by } \bigwedge_{a \notin A} \mathsf{G}_{\bar{a}} \quad \text{and}$$
$$A^{\mathrm{im}} \text{ is defined by } \bigwedge_{a \in A} (\top \mathsf{F}_a \top) \wedge \mathsf{H}_{\bar{a}}\,. \qquad \square$$

**Lemma 5** *Every unambiguous monomial $L = A_1^* a_1 \cdots A_k^* a_k A_{k+1}^\infty$ is definable in $\mathrm{ITL}^+[\mathsf{F}_a, \mathsf{L}_a, \mathsf{G}_{\bar{a}}]$.*

*Proof:* We perform an induction on $k$. For $k = 0$ we have $L = A_1^\infty$ which is definable in $\mathrm{ITL}^+[\mathsf{G}_{\bar{a}}]$ by Lemma 4. Let $k \geq 1$. Since $L$ is unambiguous, we have $\{a_1, \ldots, a_k\} \not\subseteq A_1 \cap A_{k+1}$; otherwise $(a_1 \cdots a_k)^2$ admits two different factorizations showing that $L$ is not unambiguous. First, consider the case $a_i \notin A_1$ and let $i$ be minimal with this property. Each word $\alpha \in L$ has a unique factorization $\alpha = u a_i \beta$ such that $a_i \notin \mathrm{alph}(u)$. Depending on whether the first $a_i$ of $\alpha$ coincides with the marker $a_i$ or not, we have

$$\begin{array}{lll} u \in A_1^* a_1 \cdots A_i^*, & & \beta \in A_{i+1}^* a_{i+1} \cdots A_k^* a_k A_{k+1}^\infty \quad \text{or} \\ u \in A_1^* a_1 \cdots A_j^*, & a_i \in A_j, & \beta \in A_j^* a_j \cdots A_k^* a_k A_{k+1}^\infty \end{array}$$

with $2 \leq j \leq i$. In both cases, since $L$ is unambiguous, each expression containing $u$ or $\beta$ is unambiguous. Moreover, each of these expressions is strictly shorter than $L$. By induction, for each $2 \leq j \leq k$, there exist formulas $\varphi, \psi \in \mathrm{ITL}^+[\mathsf{F}_a, \mathsf{L}_a, \mathsf{G}_{\bar{a}}]$ such that $L(\varphi) = A_1^* a_1 \cdots A_j^\infty$ and $L(\psi) = A_j^* a_j \cdots A_k^* a_k A_{k+1}^\infty$. By the above reasoning, we see that $L$ is the union of (at most $i$) languages of the form

$$\left(L(\varphi) \cap (\Gamma \setminus \{a_i\})^*\right) a_i\, L(\psi)$$

and each of them is defined by $\varphi \mathsf{F}_{a_i} \psi$.

For $a_i \notin A_{k+1}$ with $i$ maximal we consider the unique factorization $\alpha = u a_i \beta$ with $a_i \notin \mathrm{alph}(\beta)$ and again we end up with one of the two cases from above, with the difference that $1 \leq i < j \leq k$ in the second case. Inductively $L$ is defined by a disjunction of formulas $\varphi \mathsf{L}_{a_i} \psi$. $\square$

The language $L(\mathsf{H}_{\bar{a}})$ is definable in $\Pi_2 \cap \mathrm{FO}^2$ using the following formula

$$\forall x \exists y \colon \lambda(x) = a \vee \big(y > x \wedge \lambda(y) = a\big),$$

but it is not definable in $\Sigma_2$. Therefore, we have to exclude the case $r = \mathsf{H}_{\bar{a}}$ in the following lemma.

**Lemma 6** *Let $r \neq \mathsf{H}_{\bar{a}}$ be a non-empty ranker, potentially with atomic modality $\mathsf{G}_{\bar{a}}$ or $\mathsf{H}_{\bar{a}}$, then $L(r)$ is definable in $\mathrm{FO}^2 \cap \Sigma_2$.*

*Proof:* By induction on $k = |r|$ we construct formulas $\mu_r(x) \in \mathrm{FO}^2$ and $\sigma_r(x) \in \Sigma_2$ with one free variable such that for every word $\alpha \in \Gamma^\infty$ we have

$$\alpha \models \exists x \colon \mu_r(x) \quad \text{iff} \quad r(\alpha) \text{ is defined} \quad \text{iff} \quad \alpha \models \exists x \colon \sigma_r(x)$$

and if $r(\alpha)$ is defined, then

$$\alpha, x \models \mu_r(x) \qquad \text{iff} \qquad x = r(\alpha) \qquad \text{iff} \quad \alpha, x \models \sigma_r(x).$$

For $\mu_r$ we only use the variables $x$ and $y$. By interchanging the names, we can always choose whether $x$ or $y$ is the free variable of $\mu_r$. The formula $\sigma_r = \sigma_r(x_k)$ will have the form

$$\exists x_{k-1} \cdots \exists x_1 \forall y \colon \nu_r(x_k, \ldots, x_1, y).$$

For $r = \mathsf{G}_{\bar{a}}$ we set

$$\mu_r(x) \;\equiv\; \sigma_r(x) \;\equiv\; \forall y \colon \lambda(y) \neq a.$$

For $r = \mathsf{Y}_a$ we define

$$\mu_r(x) \;\equiv\; \sigma_r(x) \;\equiv\; \forall y \colon \lambda(x) = a \wedge (y \leq x \vee \lambda(y) \neq a).$$

Note that if $\alpha$ contains infinitely many $a$'s, then $\alpha \not\models \exists x \colon \mu_r(x)$ for $r = \mathsf{Y}_a$. The case $r = \mathsf{X}_a$ is symmetric; we only have to replace "$\leq$" with "$\geq$" in the above formula.

Let now $k = |r| > 1$. We first consider the formula $\mu_r$. If $r = s\,\mathsf{X}_a$, then

$$\mu_r(x) \;\equiv\; \begin{cases} \lambda(x) = a \;\wedge\; \exists y < x \colon \mu_s(y) \;\wedge \\ \forall y < x \colon \big(\lambda(y) \neq a \;\vee\; \exists x \geq y \colon \mu_s(x)\big) \end{cases}$$

saying that $x$ is an $a$-position greater than the $s$-position (in particular, $r$ is defined) and that all $a$-positions smaller than $x$ are not greater than the $s$-position. The case $r = s\,\mathsf{Y}_a$ is symmetric; we only have to replace all "$<$" by "$>$" and "$\geq$" by "$\leq$" in the above formula. If $r = s\,\mathsf{G}_{\bar{a}}$, then

$$\mu_r(x) \;\equiv\; \mu_s(x) \;\wedge\; \forall y > x \colon \lambda(y) \neq a,$$

and for $r = s\,\mathsf{H}_{\bar{a}}$ we use the same formula, but $y > x$ is replaced by $y < x$.

We now describe the construction of the formula $\sigma_r$. Suppose $r = s\,\mathsf{X}_a$ and let $\sigma_s(x_{k-1}) = \exists x_{k-2} \cdots \exists x_1 \forall y \colon \nu_s(x_{k-1}, \ldots, x_1, y)$. Then

$$\sigma_r(x_k) \;\equiv\; \begin{cases} \exists x_{k-1} \cdots \exists x_1 \forall y \colon \lambda(x_k) = a \wedge \\ \quad x_k > x_{k-1} \wedge \nu_s(x_{k-1}, \ldots, x_1, y) \wedge \\ \quad \big(y \leq x_{k-1} \vee y \geq x_k \vee \lambda(y) \neq a\big). \end{cases}$$

The semantics of $\sigma_r$ is as follows: $x_k$ is labeled by $a$ and it is strictly greater than the $s$-position $x_{k-1}$; moreover, there is no $a$-position strictly between $x_{k-1}$ and $x_k$. Hence, $x_k$ determines the $r$-position. As before, the case $r = s\,\mathsf{Y}_a$ is symmetric; we only have to replace "$>$" by "$<$" and we have to interchange "$\geq$" and "$\leq$" in the above formula. If $r = s\,\mathsf{G}_{\bar{a}}$, then

$$\sigma_r(x_k) \;\equiv\; \exists x_{k-1} \cdots \exists x_1 \forall y \colon \nu_s(x_k, x_{k-2}, \ldots, x_1, y) \wedge \big(y \leq x_k \vee \lambda(y) \neq a\big),$$

and for $r = s\,\mathsf{H}_{\bar{a}}$ we use the same formula, but $y \leq x_k$ is replaced by $y \geq x_k$. □

*Proof (Theorem 1):* We show "1 ⇒ 2 ⇒ 3 ⇒ 4 ⇒ 7 ⇒ 1" and "2 ⇒ 5 ⇒ 6 ⇒ 4".

"1 ⇒ 2": Every $\mathrm{FO}^2$-definable language is a finite union of languages of the form $P \cap A^{\mathrm{im}}$ with an unambiguous monomial $P$ and $A \subseteq \Gamma$, see [2]. Since $\mathrm{ITL}^+[\mathsf{F}_a, \mathsf{L}_a, \mathsf{G}_{\bar{a}}, \mathsf{H}_{\bar{a}}]$ is closed under finite unions and finite intersections, it suffices to show that $A^{\mathrm{im}}$ and $P$ are definable in $\mathrm{ITL}^+[\mathsf{F}_a, \mathsf{L}_a, \mathsf{G}_{\bar{a}}, \mathsf{H}_{\bar{a}}]$. This follows from Lemma 4 and Lemma 5, respectively.

"2 ⇒ 3" and "6 ⇒ 4" are trivial. "3 ⇒ 4" and "2 ⇒ 5": Proposition 1. "4 ⇒ 7" and "5 ⇒ 6": Lemma 1.

"7 ⇒ 1": By Lemma 6, for every ranker $r$ the language $L(r)$ is $\mathrm{FO}^2$-definable. Since $\mathrm{FO}^2$ is closed under Boolean operations, every ranker language is $\mathrm{FO}^2$-definable. □

## 5 The Fragment $\Sigma_2 \cap \mathrm{FO}^2$

In the following, we show that $\Sigma_2 \cap \mathrm{FO}^2$ admits characterizations in terms of eager ITL, TL, and rankers.

**Theorem 2** *Let $L \subseteq \Gamma^\infty$. The following assertions are equivalent:*

1. *$L$ is definable in $\Sigma_2$ and $\mathrm{FO}^2$.*
2. *$L$ is definable in $\mathrm{ITL}^+[\mathsf{F}_a, \mathsf{L}_a, \mathsf{G}_{\bar{a}}]$.*
3. *$L$ is definable in $\mathrm{TL}^+[\mathsf{X}_a, \mathsf{Y}_a, \mathsf{G}_{\bar{a}}]$.*
4. *$L$ is a positive ranker language with atomic modality $\mathsf{G}_{\bar{a}}$.*
5. *$L$ is a ranker language with atomic modality $\mathsf{G}_{\bar{a}}$ with the restriction that all $\mathsf{Y}$-rankers are positive.*

Note that we cannot use lazy counterparts in the above characterizations, since for example $\mathsf{Y}_a^\ell\, \mathsf{X}_a^\ell$ is defined if and only if there are infinitely many $a$'s, but this property is not $\Sigma_2$-definable.

**Lemma 7** *Let $r$ be an $\mathsf{X}$-ranker with atomic modality $\mathsf{G}_{\bar{a}}$. Then $\Gamma^\infty \setminus L(r)$ is $\Sigma_2$-definable.*

*Proof:* If $r$ is an $\mathsf{X}$-ranker which is not defined on $\alpha$, then there is a longest prefix $p$ of $r$ such that $p$ is defined on $\alpha$. Write $r = pq$. If the first modality $q$ is of the form $\mathsf{X}_a$ or $\mathsf{Y}_a$ or $\mathsf{G}_{\bar{a}}$ , then we set $s = p\,\mathsf{G}_{\bar{a}}$ or $s = p\,\mathsf{H}_{\bar{a}}$ or $s = p\,\mathsf{X}_a$, respectively. Note that if $q$ starts with $\mathsf{Y}_a$, then $p$ is a non-empty $\mathsf{X}$-ranker. In any case $s \neq \mathsf{H}_{\bar{a}}$, and therefore, $L(s)$ is $\Sigma_2$-definable by Lemma 6. Hence, we find a finite set of $\Sigma_2$-definable languages whose union is $\Gamma^\infty \setminus L(r)$. But $\Sigma_2$ is closed under union and thus $\Gamma^\infty \setminus L(r)$ is $\Sigma_2$-definable. □

*Proof (Theorem 2):* "1 ⇒ 2": A language $L$ is definable in $\Sigma_2 \cap \mathrm{FO}^2$ if and only if $L$ is a union of unambiguous monomials, see [2]. Since $\mathrm{ITL}^+[\mathsf{F}_a, \mathsf{L}_a, \mathsf{G}_{\bar{a}}]$ is closed under union, it suffices to show that every unambiguous monomial is definable in $\mathrm{ITL}^+[\mathsf{F}_a, \mathsf{L}_a, \mathsf{G}_{\bar{a}}]$; this is exactly Lemma 5.

"2 ⇒ 3": Proposition 1. "3 ⇒ 4": Lemma 1. "4 ⇒ 5": trivial.

"5 ⇒ 1": Since languages in $\Sigma_2 \cap \mathrm{FO}^2$ are closed under finite union and finite intersection, the claim follows from Lemma 6 and Lemma 7. □

Over finite words, the fragments $\mathrm{FO}^2$ and $\Delta_2$ coincide [8]. In particular, $\mathrm{FO}^2 \cap \Sigma_2 = \mathrm{FO}^2$ for finite words. Since finiteness of a word is definable in $\mathrm{FO}^2 \cap \Sigma_2$, we obtain the following corollary of Theorem 2.

**Corollary 1** *A language $L \subseteq \Gamma^*$ of finite words is definable in $\mathrm{FO}^2$ if and only if $L$ is a positive ranker language with atomic modality $\mathsf{G}_{\bar{a}}$.*

## 6 The Fragment $\Delta_2$

Over infinite words, the fragment $\Delta_2$ is a strict subclass of $\mathrm{FO}^2$. In this section, we show that $\Delta_2$ basically is $\mathrm{FO}^2$ with the lack of past formulas and past-rankers. Since eager future formulas and future-rankers coincide with their lazy counterparts, all of the characterizations in the next theorem could be replaced by their lazy pendants.

**Theorem 3** *Let $L \subseteq \Gamma^\infty$. The following assertions are equivalent:*

1. *$L$ is definable in $\Delta_2$.*
2. *$L$ is definable in $\mathrm{ITL}^+_{\mathsf{F}}[\mathsf{F}_a, \mathsf{L}_a, \mathsf{G}_{\bar{a}}]$.*
3. *$L$ is definable in $\mathrm{ITL}_{\mathsf{F}}[\mathsf{F}_a, \mathsf{L}_a]$.*
4. *$L$ is definable in $\mathrm{TL}_{\mathsf{X}}[\mathsf{X}_a, \mathsf{Y}_a]$.*
5. *$L$ is definable in $\mathrm{TL}^+_{\mathsf{X}}[\mathsf{X}_a, \mathsf{Y}_a, \mathsf{G}_{\bar{a}}]$.*
6. *$L$ is a positive $\mathsf{X}$-ranker language with atomic modality $\mathsf{G}_{\bar{a}}$.*
7. *$L$ is an $\mathsf{X}$-ranker language.*

**Lemma 8** *Every language definable in $\Delta_2$ is definable in $\mathrm{ITL}^+_{\mathsf{F}}[\mathsf{F}_a, \mathsf{L}_a, \mathsf{G}_{\bar{a}}]$.*

*Proof:* It is known that a $\Delta_2$-definable language is a finite union of unambiguous monomials $L = A_1^* a_1 \cdots A_k^* a_k A_{k+1}^\infty$ such that $\{a_j, \ldots, a_k\} \not\subseteq A_j$ for all $1 \leq j \leq k$, see [2]. Let $i$ be minimal such that $a_i \notin A_1$ and for each word $\alpha \in L$ consider the factorization $\alpha = u a_i \beta$ such that $a_i \notin \mathrm{alph}(u)$. There are two cases:

$$\begin{array}{lll} u \in A_1^* a_1 \cdots A_i^*, & & \beta \in A_{i+1}^* a_{i+1} \cdots A_k^* a_k A_{k+1}^\infty \quad \text{or} \\ u \in A_1^* a_1 \cdots A_j^*, & a_i \in A_j, & \beta \in A_j^* a_j \cdots A_k^* a_k A_{k+1}^\infty \end{array}$$

with $2 \leq j \leq i$. In each case the expression $P = A_j^* a_j \cdots A_k^* a_k A_{k+1}^\infty$ containing $\beta$ is unambiguous since $L$ is. Moreover, the expression is shorter than that for $L$ and we have $\{a_\ell, \ldots, a_k\} \not\subseteq A_\ell$ for all $j \leq \ell \leq k$. By induction $P$ is definable by an $\mathrm{ITL}^+_{\mathsf{F}}[\mathsf{F}_a, \mathsf{L}_a, \mathsf{G}_{\bar{a}}]$-formula $\psi$. By Lemma 5, we get an $\mathrm{ITL}^+[\mathsf{F}_a, \mathsf{L}_a, \mathsf{G}_{\bar{a}}]$-formula $\varphi$ defining the monomial $A_1^* a_1 \cdots A_j^\infty$. Therefore $L$ is the union of languages of the form $\varphi \, \mathsf{F}_{a_i} \, \psi$ each of which is an $\mathrm{ITL}^+_{\mathsf{F}}[\mathsf{F}_a, \mathsf{L}_a, \mathsf{G}_{\bar{a}}]$-formula by definition, given the constraints imposed on $\varphi$ and $\psi$. □

*Proof (Theorem 3):* We show "1 ⇒ 2 ⇒ 3 ⇒ 4 ⇒ 7 ⇒ 1" and "2 ⇒ 5 ⇒ 6 ⇒ 4".

"1 ⇒ 2" is Lemma 8 and "2 ⇒ 3" as well as "6 ⇒ 4" are trivial. "3 ⇒ 4" and "2 ⇒ 5": Proposition 1. "4 ⇒ 7" and "5 ⇒ 6": Lemma 1.

It remains to show "7 ⇒ 1". By Lemma 6 and Lemma 7, for every $\mathsf{X}$-ranker $r$ the languages $L(r)$ and $\Gamma^\infty \setminus L(r)$ are definable in $\Delta_2 = \Sigma_2 \cap \Pi_2$. Since $\Delta_2$ is closed under finite unions and finite intersections, the claim follows. □

## 7 The Fragment $\Pi_2 \cap \mathrm{FO}^2$

In this section we give characterizations of the fragment $\Pi_2 \cap \mathrm{FO}^2$ in term of the lazy variants of ITL, TL, and rankers. We cannot use the eager variants, since $\mathsf{Y}_a$ says that there are only finitely many $a$'s, but this property is not $\Pi_2$-definable. Also note that $\alpha, (\infty; \infty) \models \hat{\mathsf{H}}_{\bar{a}}$ for $\hat{\mathsf{H}}_{\bar{a}} = \neg(\top \, \mathsf{L}_a^\ell \, \top)$ if and only if $a \notin \mathrm{im}(\alpha)$, i.e., if and only if $a$ occurs at most finitely often. As before, this property is not $\Pi_2$-definable. This is the reason why we did not define $\mathsf{H}_{\bar{a}}^\ell$ simply as $\hat{\mathsf{H}}_{\bar{a}}$.

**Theorem 4** *Let $L \subseteq \Gamma^\infty$. The following assertions are equivalent:*

1. *$L$ is definable in $\Pi_2$ and $\mathrm{FO}^2$.*
2. *$L$ is definable in $\mathrm{ITL}^+[\mathsf{F}^\ell_a, \mathsf{L}^\ell_a, \mathsf{H}^\ell_{\bar{a}}]$.*
3. *$L$ is definable in $\mathrm{TL}^+[\mathsf{X}^\ell_a, \mathsf{Y}^\ell_a, \mathsf{H}^\ell_{\bar{a}}]$.*
4. *$L$ is a positive lazy ranker language with atomic modality $\mathsf{H}^\ell_{\bar{a}}$.*
5. *$L$ is a lazy ranker language with atomic modality $\mathsf{H}^\ell_{\bar{a}}$ with the restriction that all $\mathsf{Y}^\ell$-rankers are positive.*

**Lemma 9** *The complement $\Gamma^\infty \setminus L$ of every unambiguous monomial $L = A_1^* a_1 \cdots A_k^* a_k A_{k+1}^\infty$ is definable in $\mathrm{ITL}^+[\mathsf{F}^\ell_a, \mathsf{L}^\ell_a, \mathsf{H}^\ell_{\bar{a}}]$.*

*Proof:* We perform an induction on $k$. For $k = 0$ we have $L = A_1^\infty$ and $\Gamma^\infty \setminus A_1^\infty$ is defined by the $\mathrm{ITL}^+[\mathsf{F}^\ell_a]$-formula

$$\bigvee_{a \notin A_1} (\top \, \mathsf{F}^\ell_a \, \top).$$

Let now $k > 0$. Since $L$ is unambiguous, we have $\{a_1, \ldots, a_k\} \not\subseteq A_1 \cap A_{k+1}$; otherwise $(a_1 \cdots a_k)^2$ admits two different factorizations showing that $L$ is not unambiguous. First, consider the case $a_i \notin A_{k+1}$ and let $i$ be maximal with this property. For $\alpha \in \Gamma^\infty$ we have $\alpha \notin L$ if and only if one of the following conditions is true: The first condition is $a_i \notin \mathrm{alph}(\alpha)$ or $a_i \in \mathrm{im}(\alpha)$ and the second condition is $a_i \in \mathrm{alph}(\alpha) \setminus \mathrm{im}(\alpha)$ and the following holds for $\alpha = u a_i \beta$ with $a_i \notin \mathrm{alph}(\beta)$:

- $u \notin A_1^* a_1 \cdots A_i^\infty$ or $\beta \notin A_{i+1}^* a_{i+1} \cdots A_{k+1}^\infty$, and
- for all $i < j \leq k$ with $a_i \in A_j$ we have: $u \notin A_1^* a_1 \cdots A_j^\infty$ or $\beta \notin A_j^* a_j \cdots A_{k+1}^\infty$.

The monomials $A_1^* a_1 \cdots A_j^\infty$ for $i \leq j \leq k$ and $A_j^* a_j \cdots A_{k+1}^\infty$ for $i < j \leq k$ are unambiguous and have degree smaller than $k$. Hence by induction, there exist formulas $\varphi_j, \psi_j \in \mathrm{ITL}^+[\mathsf{F}^\ell_a, \mathsf{L}^\ell_a, \mathsf{H}^\ell_{\bar{a}}]$ such that $L(\varphi_j) = \Gamma^\infty \setminus A_1^* a_1 \cdots A_j^\infty$ and $L(\psi_j) = \Gamma^\infty \setminus A_j^* a_j \cdots A_{k+1}^\infty$. This yields the following formula for the complement of $L$:

$$\begin{aligned}&\mathsf{H}^\ell_{\bar{a}_i} \vee (\top \, \mathsf{L}^\ell_{a_i} \, (\top \, \mathsf{F}^\ell_{a_i} \, \top)) \vee \\ &\Big( ((\varphi_i \, \mathsf{L}^\ell_{a_i} \, \top) \vee (\top \, \mathsf{L}^\ell_{a_i} \, \psi_{i+1})) \;\wedge \bigwedge_{i<j\leq k, a_i \in A_j} ((\varphi_j \, \mathsf{L}^\ell_{a_i} \, \top) \vee (\top \, \mathsf{L}^\ell_{a_i} \, \psi_j)) \Big).\end{aligned}$$

The first line captures the first condition from above, since $\top \, \mathsf{L}^\ell_{a_i} \, (\top \, \mathsf{F}^\ell_{a_i} \, \top)$ is true if and only if $a_i$ appears infinitely often. Note that a term $T$ saying that $a_i$ occurs only finitely often and at least once is not required in the above formula, even though it would be natural to include it on the right-hand side of the second disjunction at the outermost level (if $T$ is false, then one of the first two terms is true). Hence, we do not have to care about the case in which the right interval of some $\mathsf{L}^\ell_{a_i}$-modality is $(\infty; \infty)$.

Let now $a_i \notin A_1$ and let $i$ be minimal with this property. For $\alpha \in \Gamma^\infty$ we have $\alpha \notin L$ if and only if one of the following conditions is true: The first condition is $a_i \notin \mathrm{alph}(\alpha)$ and the second condition is $a_i \in \mathrm{alph}(\alpha)$ and the following holds for $\alpha = u a_i \beta$ with $a_i \notin \mathrm{alph}(u)$:

- $u \notin A_1^* a_1 \cdots A_i^\infty$ or $\beta \notin A_{i+1}^* a_{i+1} \cdots A_{k+1}^\infty$, and
- for all $1 < j \leq i$ with $a_i \in A_j$ we have: $u \notin A_1^* a_1 \cdots A_j^\infty$ or $\beta \notin A_j^* a_j \cdots A_{k+1}^\infty$.

The monomials $A_1^* a_1 \cdots A_j^\infty$ for $1 < j \leq i$ and $A_j^* a_j \cdots A_{k+1}^\infty$ for $1 < j \leq i+1$ are unambiguous and have degree smaller than $k$. Hence by induction, there exist formulas $\varphi_j, \psi_j \in \mathrm{ITL}^+[\mathsf{F}_a^\ell, \mathsf{L}_a^\ell, \mathsf{H}_{\bar{a}}^\ell]$ such that $L(\varphi_j) = \Gamma^\infty \setminus A_1^* a_1 \cdots A_j^\infty$ and $L(\psi_j) = \Gamma^\infty \setminus A_j^* a_j \cdots A_{k+1}^\infty$. This yields the following formula for the complement of $L$:

$$\mathsf{H}_{\bar{a}_i}^\ell \vee \Big( \big( (\varphi_i \, \mathsf{F}_{a_i}^\ell \, \top) \vee (\top \, \mathsf{F}_{a_i}^\ell \, \psi_{i+1}) \big) \; \wedge \bigwedge_{1<j\leq i, a_i \in A_j} \big( (\varphi_j \, \mathsf{F}_{a_i}^\ell \, \top) \vee (\top \, \mathsf{F}_{a_i}^\ell \, \psi_j) \big) \Big).$$

This completes the proof. □

For the inclusion of positive lazy rankers in $\Pi_2 \cap \mathrm{FO}^2$, our proof is based on a characterization of this fragment in terms of the *alphabetic topology* over finite and infinite words [2]. A base of the open subsets of this topology is given by the sets of the form $uA^\infty$ for $u \in \Gamma^*$ and $A \subseteq \Gamma$. A language is *closed* if its complement is open. The *closure* $\overline{L}$ of a language is the intersection of all closed sets containing $L$. A word $\alpha$ belongs to $\overline{L}$ if for every finite prefix $u$ of $\alpha$ there exists $\gamma \in A^\infty$ for $A = \mathrm{im}(\alpha)$ such that $u\gamma \in L$. A language $L$ is closed if and only if $\overline{L} \subseteq L$.

**Lemma 10** *Let $L \subseteq \Gamma^\infty$. If $L$ is a lazy ranker language with atomic modality $\mathsf{H}_{\bar{a}}^\ell$ such that all $\mathsf{Y}^\ell$-rankers are positive, then $L$ is closed in the alphabetic topology.*

*Proof:* A lazy ranker starting with a future modality is equivalent to its eager counterpart. For pure eager $\mathsf{X}$-rankers $r$ we have shown in Lemma 6 and Lemma 7 that both $L(r)$ and $\Gamma^\infty \setminus L(r)$ are $\Sigma_2$-definable, and hence, $L(r)$ and $\Gamma^\infty \setminus L(r)$ are open in the alphabetic topology, i.e., $L(r)$ is clopen. For every $\mathsf{X}$-ranker $r$ we have $L(r\,\mathsf{H}_{\bar{a}}) = L(r) \setminus L(r\,\mathsf{Y}_a)$. Therefore, every lazy $\mathsf{X}^\ell$-ranker $r$ (possibly with atomic $\mathsf{H}_{\bar{a}}^\ell$-modality) generates a clopen language $L(r)$.

It remains to show that $L(r)$ is closed for every $\mathsf{Y}^\ell$-ranker $r$. The ranker $r$ may end with $\mathsf{H}_{\bar{a}}^\ell$. We show that the closure of $L(r)$ in the alphabetic topology is contained in $L(r)$. Suppose $\alpha \in \overline{L(r)}$ and let $A = \mathrm{im}(\alpha)$. Let $s$ be the maximal pure prefix of $r$, i.e., $r \in s\,\{\varepsilon, \mathsf{H}_{\bar{a}}^\ell \mid a \in \Gamma\}$, and let $k = |s| + 1$. Write $\alpha = uv_1 \cdots v_k \beta$ with $\mathrm{alph}(v_i) = A$ and $\beta \in A^\infty \cap A^{\mathrm{im}}$. Since $\alpha \in \overline{L(r)}$, there exists $\gamma \in A^\infty$ such that $uv_1 \cdots v_k \gamma \in L(r)$, i.e., $r$ is defined on the word $\alpha' = uv_1 \cdots v_k \gamma$. If $s(\alpha') = \infty$, then $s(\alpha) = \infty$, since $\mathrm{im}(\alpha') \subseteq A = \mathrm{im}(\alpha)$. Moreover, $r(\alpha)$ is defined, since $\mathrm{alph}(\alpha') = \mathrm{alph}(uv_1) = \mathrm{alph}(\alpha)$. Let now $s(\alpha') \neq \infty$. We have to distinguish two cases.

The first case is that all letters occurring in $s$ are from $A$. Then $s(\alpha) = \infty$ (in particular $s(\alpha)$ is defined) and $s(\alpha') > |uv_1|$ by choice of $k$. This shows, that $r(\alpha)$ is defined if $r = s$. Now, if $r = s\,\mathsf{H}_{\bar{a}}^\ell$, then $a \notin \mathrm{alph}(uv_1) = \mathrm{alph}(\alpha)$, and hence $r(\alpha)$ is defined.

The second case is $s = s_1\,\mathsf{Y}_b^\ell\,s_2$ such that $b \notin A$ and all letter from $s_1$ are in $A$. Note that we cannot have the situation $s = s_1\,\mathsf{X}_b^\ell\,s_2$ with $b \notin A$ and all letter from $s_1$ are in $A$, since then $s$ would be undefined on $\alpha'$. Then $s_1\,\mathsf{Y}_b^\ell(\alpha') = s_1\,\mathsf{Y}_b^\ell(\alpha) \leq |u|$. Again, by choice of $k$, it follows that $s_1\,\mathsf{Y}_b^\ell\,s_2(\alpha') = s_1\,\mathsf{Y}_b^\ell\,s_2(\alpha)$. Therefore, even if $r$ ends with $\mathsf{H}_{\bar{a}}^\ell$, we see that $r(\alpha)$ is defined.

In any case, we have $\alpha \in L(r)$. This completes the proof. □

*Proof (Theorem 4):* "1 ⇒ 2" A language is in $\Pi_2 \cap \mathrm{FO}^2$ if and only if it is the intersection of complements of unambiguous monomials, see [2]. By Lemma 9, such complements are definable in $\mathrm{ITL}^+[\mathsf{F}_a^\ell, \mathsf{L}_a^\ell, \mathsf{H}_{\bar{a}}^\ell]$. Since $\mathrm{ITL}^+[\mathsf{F}_a^\ell, \mathsf{L}_a^\ell, \mathsf{H}_{\bar{a}}^\ell]$ is closed under intersection, the claim follows.

"2 ⇒ 3" is Proposition 2 and "3 ⇒ 4" follows from Lemma 1. "4 ⇒ 5" is trivial.

"5 ⇒ 1": It is easy to see that any lazy ranker language is definable in $\mathrm{TL}[\mathsf{X}_a, \mathsf{Y}_a]$. Hence, by Theorem 1, any such language is in $\mathrm{FO}^2$. Closed languages are closed under finite union and intersection. Therefore, by Lemma 10, lazy ranker languages with atomic modality $\mathsf{H}_{\bar{a}}^\ell$ and with only positive $\mathsf{Y}^\ell$-rankers are closed in the alphabetic topology. Languages which are closed in the alphabetic topology and which are $\mathrm{FO}^2$-definable are in $\Pi_2$, see [2]. □

For completeness, we give a counterpart of Theorem 1 using the lazy versions ITL, TL, and rankers.

**Theorem 5** *For $L \subseteq \Gamma^\infty$ the following assertions are equivalent:*

1. *$L$ is definable in* $\mathrm{FO}^2$.
2. *$L$ is definable in* $\mathrm{ITL}^+[\mathsf{F}^\ell_a, \mathsf{L}^\ell_a, \mathsf{G}^\ell_{\bar{a}}, \mathsf{H}^\ell_{\bar{a}}]$.
3. *$L$ is definable in* $\mathrm{ITL}[\mathsf{F}^\ell_a, \mathsf{L}^\ell_a]$.
4. *$L$ is definable in* $\mathrm{TL}[\mathsf{X}^\ell_a, \mathsf{Y}^\ell_a]$.
5. *$L$ is definable in* $\mathrm{TL}^+[\mathsf{X}^\ell_a, \mathsf{Y}^\ell_a, \mathsf{G}^\ell_{\bar{a}}, \mathsf{H}^\ell_{\bar{a}}]$.
6. *$L$ is a positive ranker language with atomic modalities $\mathsf{G}^\ell_{\bar{a}}$ and $\mathsf{H}^\ell_{\bar{a}}$.*
7. *$L$ is a lazy ranker language.*

*Proof:* "1 ⇒ 2": Every $\mathrm{FO}^2$-definable language is a finite intersection of languages of the form $\Gamma^\infty \setminus (P \cap A^{\mathrm{im}}) = (\Gamma^\infty \setminus P) \cup (\Gamma^\infty \setminus A^{\mathrm{im}})$ with an unambiguous monomial $P$ and $A \subseteq \Gamma$, see [2]. Since $\mathrm{ITL}^+[\mathsf{F}^\ell_a, \mathsf{L}^\ell_a, \mathsf{G}^\ell_{\bar{a}}, \mathsf{H}^\ell_{\bar{a}}]$ is closed under finite unions and finite intersections, it suffices to show that $\Gamma^\infty \setminus P$ and $\Gamma^\infty \setminus A^{\mathrm{im}}$ are definable in $\mathrm{ITL}^+[\mathsf{F}^\ell_a, \mathsf{L}^\ell_a, \mathsf{G}^\ell_{\bar{a}}, \mathsf{H}^\ell_{\bar{a}}]$. For $\Gamma^\infty \setminus P$, this is shown in Lemma 9 and $\Gamma^\infty \setminus A^{\mathrm{im}}$ is defined by

$$\bigvee_{b \notin A} (\top \, \mathsf{L}^\ell_b \, (\top \, \mathsf{X}^\ell_b \, \top)) \;\vee\; \bigvee_{b \in A} (\top \, \mathsf{L}^\ell_b \, \mathsf{G}^\ell_{\bar{b}}).$$

"2 ⇒ 3" is trivial.

"3 ⇒ 4": Proposition 2.

"4 ⇒ 5": Using a similar approach as in Lemma 1 we can remove all negations. For this, we apply De Morgan's laws and the following rules for moving negations to the innermost level:

$$\begin{aligned} \neg \, \mathsf{X}^\ell_a \, \varphi &\equiv \mathsf{G}^\ell_{\bar{a}} \vee \mathsf{X}^\ell_a \, \neg\varphi, \\ \neg \, \mathsf{Y}^\ell_a \, \varphi &\equiv \mathsf{H}^\ell_{\bar{a}} \vee \mathsf{Y}^\ell_a \, \neg\varphi. \end{aligned}$$

Note that $\mathsf{X}^\ell_a \bot \;\equiv\; \bot \;\equiv\; \mathsf{Y}^\ell_a \bot$. After incorporating all constants, we end up with an equivalent formula in $\mathrm{TL}^+[\mathsf{X}^\ell_a, \mathsf{Y}^\ell_a, \mathsf{G}^\ell_{\bar{a}}, \mathsf{H}^\ell_{\bar{a}}]$.

"5 ⇒ 6" is proved in Lemma 1 (even though not stated explicitly due to lack of space).

"6 ⇒ 7" is trivial.

"7 ⇒ 1": Every lazy $\mathsf{X}$-ranker is equivalent to its eager counterpart and hence, it generates a $\mathrm{TL}[\mathsf{X}_a, \mathsf{Y}_a]$-definable language. Let now $r = \mathsf{Y}^\ell_{a_1} \mathsf{Z}^\ell_{a_2} \cdots \mathsf{Z}^\ell_{a_k}$ with each $\mathsf{Z}^\ell_{a_i}$ being either $\mathsf{X}^\ell_{a_i}$ or $\mathsf{Y}^\ell_{a_i}$. For $A \subseteq \Gamma$ we define the following macro:

$$A \subseteq \mathrm{im} \;\equiv\; \bigwedge_{a \in A} (\mathsf{X}_a \top \wedge \neg \mathsf{Y}_a \top)$$

Now, $L(r)$ is defined by

$$\bigvee_{\substack{i = 0, \ldots, k \\ \mathsf{Z}^\ell_{a_{i+1}} \neq \mathsf{X}^\ell_{a_{i+1}}}} \Big( \{a_1, \ldots, a_i\} \subseteq \mathrm{im} \;\wedge\; \mathsf{Z}^\ell_{a_{i+1}} \cdots \mathsf{Z}^\ell_{a_k} \top \Big)$$

Therefore, every lazy ranker language is $\mathrm{TL}[\mathsf{X}_a, \mathsf{Y}_a]$-definable, and by Theorem 1 it is $\mathrm{FO}^2$-definable. □

# 8 Conclusion

We have given an eager and a lazy generalization of rankers for infinite words. Together with the usual rankers over finite words, we obtained combinatorial descriptions of various fragments of first-order logic $\mathrm{FO}[<]$ over finite and infinite words. Without negation the eager variant cannot express that there are infinitely many occurrences of some letter. This leads to a characterization of the fragment $\Sigma_2 \cap \mathrm{FO}^2$. Similarly, we cannot say that some letter occurs only finitely often in the lazy version, and this yields $\Pi_2 \cap \mathrm{FO}^2$. Both eager and lazy rankers are suitable for describing $\mathrm{FO}^2$ and $\Delta_2$. Intermediate steps in all our proofs have been unambiguous ITL and unambiguous TL — both in an eager and a lazy variant. The following table summarizes some characterizations of the fragments. For conciseness we introduce some additional terminology. By $\mathrm{Rankers}[\mathcal{C}]$ we denote the class of ranker languages with atomic modalities $\mathcal{C}$. If $\mathcal{C}$ is empty we simply write Rankers. The positive fragment is denoted by $\mathrm{Rankers}^+[\mathcal{C}]$ and $\mathrm{Rankers}_{\mathsf{X}}[\mathcal{C}]$ are the $\mathsf{X}$-ranker languages. If we prepend an $\ell$ we mean the respective lazy pendant.

| FO-Logic | Interval Logic | Temporal Logic | Rankers | |
|---|---|---|---|---|
| $\mathrm{FO}^2$ | $\mathrm{ITL}[\mathsf{F}_a, \mathsf{L}_a]$<br>$\mathrm{ITL}[\mathsf{F}_a^\ell, \mathsf{L}_a^\ell]$<br>$\mathrm{ITL}^+[\mathsf{F}_a, \mathsf{L}_a, \mathsf{G}_{\bar{a}}, \mathsf{H}_{\bar{a}}]$<br>$\mathrm{ITL}^+[\mathsf{F}_a^\ell, \mathsf{L}_a^\ell, \mathsf{G}_{\bar{a}}^\ell, \mathsf{H}_{\bar{a}}^\ell]$ | $\mathrm{TL}[\mathsf{X}_a, \mathsf{Y}_a]$<br>$\mathrm{TL}[\mathsf{X}_a^\ell, \mathsf{Y}_a^\ell]$<br>$\mathrm{TL}^+[\mathsf{X}_a, \mathsf{Y}_a, \mathsf{G}_{\bar{a}}, \mathsf{H}_{\bar{a}}]$<br>$\mathrm{TL}^+[\mathsf{X}_a^\ell, \mathsf{Y}_a^\ell, \mathsf{G}_{\bar{a}}^\ell, \mathsf{H}_{\bar{a}}^\ell]$ | Rankers<br>$\ell$-Rankers<br>$\mathrm{Rankers}^+[\mathsf{G}_{\bar{a}}, \mathsf{H}_{\bar{a}}]$<br>$\ell\text{-}\mathrm{Rankers}^+[\mathsf{G}_{\bar{a}}^\ell, \mathsf{H}_{\bar{a}}^\ell]$ | Thm. 1<br>Thm. 5 |
| $\Sigma_2 \cap \mathrm{FO}^2$ | $\mathrm{ITL}^+[\mathsf{F}_a, \mathsf{L}_a, \mathsf{G}_{\bar{a}}]$ | $\mathrm{TL}^+[\mathsf{X}_a, \mathsf{Y}_a, \mathsf{G}_{\bar{a}}]$ | $\mathrm{Rankers}^+[\mathsf{G}_{\bar{a}}]$ | Thm. 2 |
| $\Pi_2 \cap \mathrm{FO}^2$ | $\mathrm{ITL}^+[\mathsf{F}_a^\ell, \mathsf{L}_a^\ell, \mathsf{H}_{\bar{a}}^\ell]$ | $\mathrm{TL}^+[\mathsf{X}_a^\ell, \mathsf{Y}_a^\ell, \mathsf{H}_{\bar{a}}^\ell]$ | $\ell\text{-}\mathrm{Rankers}^+[\mathsf{H}_{\bar{a}}^\ell]$ | Thm. 4 |
| $\Delta_2$ | $\mathrm{ITL}_{\mathsf{F}}[\mathsf{F}_a, \mathsf{L}_a]$<br>$\mathrm{ITL}_{\mathsf{F}}[\mathsf{F}_a^\ell, \mathsf{L}_a^\ell]$<br>$\mathrm{ITL}_{\mathsf{F}}^+[\mathsf{F}_a, \mathsf{L}_a, \mathsf{G}_{\bar{a}}]$<br>$\mathrm{ITL}_{\mathsf{F}}^+[\mathsf{F}_a^\ell, \mathsf{L}_a^\ell, \mathsf{G}_{\bar{a}}^\ell]$ | $\mathrm{TL}_{\mathsf{X}}[\mathsf{X}_a, \mathsf{Y}_a]$<br>$\mathrm{TL}_{\mathsf{X}}[\mathsf{X}_a^\ell, \mathsf{Y}_a^\ell]$<br>$\mathrm{TL}_{\mathsf{X}}^+[\mathsf{X}_a, \mathsf{Y}_a, \mathsf{G}_{\bar{a}}]$<br>$\mathrm{TL}_{\mathsf{X}}^+[\mathsf{X}_a^\ell, \mathsf{Y}_a^\ell, \mathsf{G}_{\bar{a}}^\ell]$ | $\mathrm{Rankers}_{\mathsf{X}}$<br>$\ell\text{-}\mathrm{Rankers}_{\mathsf{X}}$<br>$\mathrm{Rankers}_{\mathsf{X}}^+[\mathsf{G}_{\bar{a}}]$<br>$\ell\text{-}\mathrm{Rankers}_{\mathsf{X}}^+[\mathsf{G}_{\bar{a}}^\ell]$ | Thm. 3 |

## Open Problems

Rankers over finite words have been introduced for characterizing quantifier alternation within $\mathrm{FO}^2$. We conjecture that similar results for infinite words can be obtained using our generalizations of rankers.

Over infinite words, the class of $\mathsf{X}$-ranker languages correspond to the fragment $\Delta_2$. Over finite words however, $\mathsf{X}$-ranker languages form a strict subclass of $\Delta_2$ (which for finite words coincides with $\mathrm{FO}^2$). An algebraic counterpart is still missing. The main problem is that $\mathsf{X}$-ranker languages do not form a variety of languages.

A well-known theorem by Schützenberger [5] implies that over finite words, arbitrary finite unions of unambiguous monomials and finite *disjoint* unions of unambiguous monomials describe the same class of languages. In the case of infinite words, it is open whether one can require that unambiguous polynomials are disjoint unions of unambiguous monomials without changing the class of languages.

**Acknowledgments.** We thank Volker Diekert for a suggestion which led to Theorem 2. We also thank the anonymous referees for several useful suggestions which helped to improve the presentation of the paper.

## References

[1] Diekert, V., Gastin, P., Kufleitner, M.: *A survey on small fragments of first-order logic over finite words.* International Journal of Foundations of Computer Science 19(3), 513–548 (June 2008), special issue DLT 2007

[2] Diekert, V., Kufleitner, M.: *Fragments of first-order logic over infinite words.* STACS 2009. Dagstuhl Seminar Proceedings, vol. 09001, pp. 325–336 (2009)

[3] Kufleitner, M.: *Polynomials, fragments of temporal logic and the variety* DA *over traces.* Theoretical Computer Science 376, 89–100 (2007), special issue DLT 2006

[4] Lodaya, K., Pandya, P.K., Shah, S.S.: *Marking the chops: an unambiguous temporal logic.* IFIP TCS 273, 461–476 (2008)

[5] Schützenberger, M.P.: *Sur le produit de concaténation non ambigu.* Semigroup Forum 13, 47–75 (1976)

[6] Schwentick, T., Thérien, D., Vollmer, H.: *Partially-ordered two-way automata: A new characterization of* DA. Proceedings of DLT'01, LNCS vol. 2295, pp. 239–250. Springer (2001)

[7] Tesson, P., Thérien, D.: *Diamonds are forever: The variety* DA. Semigroups, Algorithms, Automata and Languages, 475–500 (2002)

[8] Thérien, D., Wilke, Th.: *Over words, two variables are as powerful as one quantifier alternation.* Proceedings of STOC'98, 234-240 (1998). (1998)

[9] Weis, Ph., Immerman, N.: *Structure theorem and strict alternation hierarchy for* $\mathrm{FO}^2$ *on words.* Logical Methods in Computer Science 5(3:3), 1–23 (2009)